\documentclass[twocolumn]{emulateapj}  
\usepackage[left]{lineno}
\usepackage{epstopdf}
\usepackage{ulem}
\usepackage{amssymb}
\usepackage{natbib}
\usepackage{times}
\usepackage{graphicx}
\usepackage[usenames,dvipsnames]{pstricks}
\usepackage{psfrag}
\usepackage{lipsum}
\usepackage{epstopdf}
\usepackage[colorlinks=true,linkcolor=blue,citecolor=blue]{hyperref}

\shorttitle{High-redshift short GRBs}
\shortauthors{Dichiara et al.}

\begin{document}

\title{Evidence of extended emission in GRB 181123B and other high-redshift short GRBs}


\author{
S. Dichiara\altaffilmark{1,2}\email{ dichiara@umd.edu},
E. Troja\altaffilmark{1,2},
P. Beniamini\altaffilmark{3,4},
B. O'Connor\altaffilmark{1,2,5,6},
M. Moss\altaffilmark{5,2},
A. Y. Lien\altaffilmark{2,7},
R. Ricci\altaffilmark{8,9},
L. Amati\altaffilmark{10},
G. Ryan\altaffilmark{1,2},
T. Sakamoto\altaffilmark{11}
}
\affil{$^{1}$ Department of Astronomy, University of Maryland, College Park, MD 20742-4111, USA \\
$^{2}$ Astrophysics Science 
Division, NASA Goddard Space Flight Center, 8800 Greenbelt Rd, Greenbelt, MD 20771, USA\\
$^{3}$ Division of Physics, Mathematics and Astronomy, California Institute of Technology, Pasadena, CA 91125, USA\\
$^{4}$Astrophysics Research Center of the Open University (ARCO), The Open University of Israel, P.O Box 808, Ra’anana 43537, Israel\\
$^{5}$ Department of Physics, The George Washington University, 725 21st Street NW, Washington, DC 20052, USA\\
$^{6}$ Astronomy, Physics and Statistics Institute of Sciences (APSIS)\\
$^{7}$ Department of Physics, University of Maryland, Baltimore County, 1000 Hilltop Circle, Baltimore, MD 21250, USA \\
$^{8}$ Istituto Nazionale di Ricerche Metrologiche - Torino, Italy \\
$^{9}$ INAF - Istituto di Radioastronomia - Bologna, Italy \\
$^{10}$ INAF – Osservatorio di Astrofisica e Scienza dello Spazio di Bologna, Via P. Gobetti 101, I-40129 Bologna, Italy \\
$^{11}$ Department of Physics and Mathematics, 
Aoyama Gakuin University, 5-10-1 Fuchinobe, Chuoku, Sagamiharashi Kanagawa 252-5258, Japan 
}

\begin{abstract}
We study the high-energy properties of GRB 181123B, a short gamma-ray burst (sGRB) at redshift $z$\,$\approx$1.75. We show that, despite its nominal short duration with $T_{90}$\,$<$\,2 s, this burst displays evidence of a temporally extended emission (EE) at high energies and that the same trend is observed in the majority of sGRBs at $z$\,$\gtrsim$1. 
We discuss the impact of instrumental selection effects on the GRB classification, stressing that the measured $T_{90}$ is not an unambiguous indicator of the burst physical origin. By examining their environment (e.g. stellar mass, star formation, offset distribution), we find that these high-$z$ sGRBs share many properties of long GRBs at a similar distance and are consistent with a short-lived progenitor system. 
If produced by compact binary mergers, these sGRBs with EE may 
be easier to localize at large distances and
herald a larger population of sGRBs in the early universe. 

\end{abstract}

\keywords{gamma-ray burst: general -- star: neutron -- nuclear reactions, nucleosynthesis, abundances -- gravitational waves}

\section{Introduction}
\label{sec:intro}

Gamma-ray busts (GRBs) are brief flashes of gamma-ray radiation detected at a rate of $\sim$ 1 per day.
They are grouped  into two main classes based on their bimodal distribution in duration \citep{Kouveliotou1993}: long duration bursts 
are related to the collapse of very massive stars \citep[e.g. ][]{Woosley2006}, 
whereas short duration GRBs (sGRBs) are traditionally connected to mergers of compact objects \citep[e.g. ][]{Eichler1989, Narayan1992}. 
The first joint detection of a gravitational wave (GW) event (GW170817) and a sGRB (GRB170817A; \citealt{Abbott2017a}) unambiguously established the link between neutron star (NS) mergers and some sGRBs. The subsequent discovery of the kilonova AT2017gfo provided the first robust evidence for production of heavy metals in the merger ejecta \citep[e.g.,][]{Watson2019}, thus confirming that NS mergers are one of the astrophysical sites of r-process nucleosynthesis. 
This is also supported by the detection of candidate kilonovae in some nearby sGRBs \citep[e.g.][]{Tanvir2013,Troja2019,Ascenzi2019, Jin2020}. 
However, the cosmic origin of r-process elements is still far from being settled \citep{Cowan2020}. Many open questions remain, among them is 
whether NS mergers can account for the r-process enhancement of metal-poor stars and dwarf galaxies \citep{Roederer2016,Beniamini2016,Beniamini2016a,Frebel2018,Skuladottir2018}.

While the gravitational wave signal from these mergers can be detected out to a few hundred Mpc \citep{Abbott2020},
sGRBs span a wider range of redshifts \citep[from $z$\,$\sim$0.1 to $z$\,$\gtrsim$2; ][]{Selsing2019}, and therefore are a unique tool to pinpoint NS mergers across all cosmic times. 
The redshift distribution of sGRBs is a key observational input to infer the age of their stellar progenitors \citep{McCarthy2020,Anand2018,Behroozi2014,Zheng2007}, and thus estimate their contribution to the cosmic chemical evolution. 
For instance,  sGRBs can be used to infer 
the observational delay time\footnote{The delay time is defined as the time elapsed between the last
burst of star formation and the merger.}
distribution (DTD) of NS mergers, which can then be compared to the theoretically predicted DTDs for different formation channels.
The DTD implied by sGRB observations is consistent with
either a log-normal  \citep[e.g.][]{Paterson2020} or a power-law distribution 
with index between -1, in agreement with population studies \citep{Dominik2012}, and -1.5 \citep{DAvanzo2014}.
The shallower slope ($\sim$-1) points to a population of old mergers, which underpredicts the r-process abundances in early metal-poor stars \citep{Hotokezaka2018} and is inconsistent with observations of Galactic binary neutron stars (BNS) which imply typical delay times shorter than 1 Gyr \citep{BP19}. Such a shallow DTD may require invoking another prompt channel of r-process production, such as supernovae \citep[e.g. ][]{Fryer2006}, with similar overall yields.
Instead, a steeper slope of the DTD ($\sim$-1.5) implies the existence of tighter binary systems, which merge on timescales of $<100$ Myr \citep{Belczynski2006}. This prompt channel of mergers would yield a better agreement with the Galactic chemical composition \citep{Cote2019}. 

To date, out of $\approx$130 sGRBs detected by the \textit{Neil Gehrels Swift} observatory \citep{gehrels04}, only 25\% have a measured redshift, and less than 5\% are found at $z>$1. 
Any inference about the DTD of sGRB progenitors is therefore affected by large uncertainties due to the small number of events and to complex observational biases.
These biases affect the galaxy’s identification, its redshift measurement, and the classification of the GRB itself.
In particular, the $T_{90}$\footnote{The time during which the cumulative time counts increase from 5 to 95\% above background, thus encompassing 90\% of the total GRB counts \citep{Kouveliotou1993}.}, largely used to parametrize the burst duration and discriminate between the two classes of bursts, can be significantly affected by instrumental selection effects, especially for high redshift GRBs.
Several works already explored the limitations of an empirical GRB classification and proposed new methods \citep{Dainotti2020, Dereli2020, Jespersen2020,Li2020, Bromberg2013,Virgili2011,Bloom2008,Zhang2009}, yet the identification of a sGRB remains strongly connected with its reported $T_{90}$. 

Ambiguity in the classification of GRBs arise not only between the two main classes of bursts (long versus short), but also
 in the identification of ``hybrid'' events, such as short GRBs with extended emission (sGRBEE; \citealt{Norris2006, Gehrels2006}). 
These bursts are characterized by a main peak with the typical features of a sGRB (short duration, hard spectrum and negligible spectral lag), followed by a lull and then a temporally extended tail, which is spectrally softer and can last several tens of seconds. 
These sGRBEEs are found in a heterogeneous environment and are not associated with any bright SNe \citep[e.g.][]{Barthelmy2005a,Covino2006,DAvanzo2009}, although 
constraints are available only for a few nearby events. 
This evidence seems to favor a physical link between sGRBEEs and sGRBs, although the long duration ($T_{90}$ $\gg$ 2 s) of their high-energy emission poses a challenge to our common understanding of NS mergers.  Possible explanations for their phenomenology include the formation of a long-lived highly magnetized NS \citep[magnetar;][]{Gompertz2013}, a NS-black hole (BH) encounter \citep{Troja2008}, a core-collapse fallback SN \citep{Valenti2009}, or more simply viewing angle effects \citep{BarkovPozanenko2011,Oganesyan2020}.
The latter scenario is consistent with the redshift distribution found by \cite{Anand2018} who observe no significant difference between the two classes
and support an old progenitor system.
Alternatively, sGRBEEs may not fit at all into the collapsar/merger dichotomy and herald a novel
and rare channel of GRB production \citep[e.g.][]{lyutikov17,king07,fryer99}. 
Therefore it is still an open question whether the diverse phenomenology of the high-energy emission reflects a true diversity in progenitors or central engines, and whether sGRBEEs could trace the evolution of NS mergers as sGRBs do. 

The distinction between the two sub-classes of bursts (sGRBs and sGRBEEs) is generally evident at low redshifts. With typical luminosities in the range $L_X$\,$\approx10^{49}$--$10^{50}$ erg/s, the EE is readily detected by \textit{Swift} and drives the GRB duration to $\sim$\,100~s. In these cases, the main observational challenge is to discriminate between sGRBEEs and standard long duration GRBs \citep{Gehrels2006}. However, at higher redshifts ($z$\,$\gtrsim$\,1)  - which are critical to probe the DTD of NS mergers - instrumental effects become important and the EE more easily escapes detection, blurring the distinction between sGRBs and sGRBEEs. 

In this work, we present our study of the high-energy properties of the short GRB 181123B, recently localized by \citet{Paterson2020} at redshift $\approx$1.75, and other candidate high-$z$ sGRBs. 
We show that, despite the canonical classification of a $T_{90}<$2 s, the majority of these bursts show evidence of long-lived emission at high energies.  
The paper is organized as follows: in Section~\ref{sec:data} we report the procedure used to analyze the \textit{Swift} data of GRB 181123B and of a selected sample of short GRBs at high redshift. In Section~\ref{sec:results} we report the results obtained from this analysis, studying the possible presence of an extended emission during the prompt phase. We used simulations to investigate how instrumental selection effects can affect the burst classification and we studied the host galaxy properties deriving information about the burst environment.
In Section~\ref{sec:discussion} we discuss the implications of our results, considering different scenarios for the burst classification and comparing them with the observed DTD. 
Conclusions are summarized in Section~\ref{sec:conclusions}. Uncertainties are quoted at the 68\% confidence level (1~$\sigma$) for each parameter of interest and upper limits are given at a 3~$\sigma$ level, unless otherwise stated. We adopt a standard $\Lambda$CDM cosmology \citep{planck2020}.

\section{Observations and data analysis}
\label{sec:data}

\subsection{\textit{Swift}/BAT}
\label{sec:bat_analysis}

GRB 181123B triggered  the  Burst  Alert  Telescope  \citep[BAT;][]{Barthelmy2005b} aboard {\it Swift} at  $T_{0}$ = 05:33:03 UT on November 23, 2018.  The spacecraft rapidly slewed to the burst position in order to begin observations with its narrow field instruments.

The \textit{Swift}/BAT data were processed using the \textsc{heasoft} package (v6.25). The energy calibration was applied with \textsc{bateconvert} and the mask weighting was included with \textsc{batmaskwtevt}. 
We derive a $T_{90}=0.26 \pm 0.05$~s running \textsc{battblock} over the 16~ms mask-weighted light curve extracted in the 15-350 keV energy band.
The time-averaged spectrum, extracted over the time interval of $T_{0}$+0.032~s and $T_{0}$+0.312~s, is described by a simple power law with photon index $\Gamma_{\gamma}$=0.72$^{+0.17}_{-0.16}$, in agreement with the online BAT catalog\footnote{https://swift.gsfc.nasa.gov/results/batgrbcat/} \citep{Lien2016}.
According to this model, the fluence is $(1.3 \pm 0.2) \times 10^{-7}$ erg cm$^{-2}$ in the 15-150 keV energy band (observer frame).
At a redshift $z$\,$\sim$1.75, it
corresponds to an isotropic-equivalent energy of $E_{\gamma,iso} = (2.6 \pm 0.6) \times 10^{50}$ erg (15-150 keV; rest frame).

\begin{figure}
\begin{center}
\includegraphics[width=0.95\columnwidth]{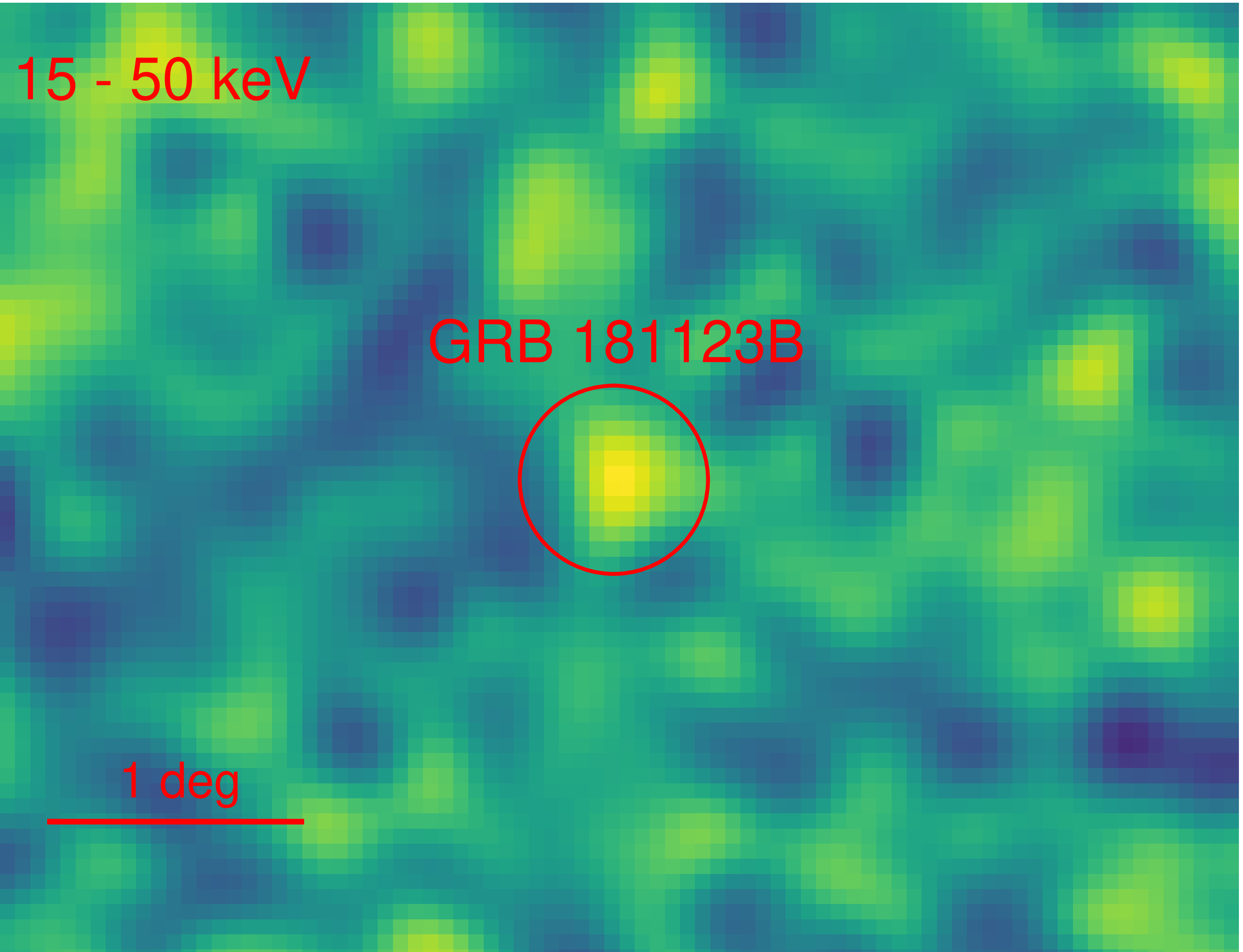}
\end{center}
\caption{BAT image of GRB 181123B obtained using the procedure described in Section~\ref{sec:bat_analysis}.}
\label{bat_181123B_image}
\end{figure}

A preliminary analysis of the BAT data reported a marginal ($\approx$3\,$\sigma$) evidence of temporally EE following the main pulse \citep{GCN.23443}. 
However, standard BAT tools are optimized for pointed observations, and assume that the source remains at a fixed position on the detector plane. This is not valid for slewing intervals, occurring in this case between 16~s and 64~s post-trigger. If the movement of the source across the detector is not properly taken into account, signal from the EE might be underestimated. 
In order to determine the presence of EE in this burst, we follow the  procedure outlined by \cite{Copete2012} for the analysis of BAT slew data. First, we accumulate the event-mode data into a Detector Plane Image (DPI) using a fine time bin of 0.2~s, during which the source position can be considered constant. After screening for bad pixels with the task \textsc{bathotpix}, we use the \textsc{batfftimage} to create a sky image and apply standard corrections for geometrical effects and flat-fielding.
At this step, we also increase the sampling of the point spread function (PSF) from the default value of 2 to 4. 
Images collected from $T_{0}$+2 s to $T_{0}$+150 s are then coadded by variance-weighted addition, and the significance of the source is determined with the task \textsc{batcelldetect}. We perform a targeted search at the GRB position allowing to fit for the peak position within a window of 2 pixels. 
We restrict our analysis to the 15-50~keV energy band, as the EE is generally more prominent at lower energies \citep{Norris2006}. The signal accumulated in this soft tail reach a significance of about 4~$\sigma$, and is shown in Figure~\ref{bat_181123B_image}. 
Previous studies of BAT images \citep{Troja2010} found that the significance calculated by 
\textsc{batcelldetect} roughly corresponds to the probability value of a Gaussian distribution. In this case the chance of a spurious detection is $\lesssim$10$^{-4}$.
Integrating this signal we found a fluence of $\approx$ $10^{-7}$ erg cm$^{-2}$ (15-50 keV), corresponding to an energy of $\approx$ 4$\times 10^{50}$ erg (15-50 keV; rest frame).

\subsection{\textit{Swift}/XRT}
\label{sec:xrt_analysis}

Observations with the X-Ray Telescope \citep[XRT; ][]{Burrows2005} aboard {\it Swift} began at $T_{0}$+80.2~s 
and continued until $\sim$0.7 days after the trigger, when the afterglow faded below detection threshold. 

We  derive  the count rate X-ray light curve  (0.3-10~keV) and the relevant spectra from the UK Swift Data Centre repository\footnote{https://www.swift.ac.uk/xrt\_products/}
\citep{Evans2007,Evans2009}. 
The X-ray temporal behavior can be described as a simple power-law, $F \propto t^{-\alpha}$, with decay index $\alpha$~=~1.40$\pm$0.08.
The time average spectrum integrated between 95~s and 16~ks 
is well fit with a simple power law with a photon index of $\Gamma_{X}=2.0_{-0.2}^{+0.3}$, a galactic absorption 
$N_H$ = 3.1 $\times$ 10$^{20}$ cm$^{-2}$ \citep{Willingale2013} and an intrinsic absorption $N_{H,int} <$ 9 $\times$ 10$^{20}$ cm$^{-2}$.
We use these spectral parameters to convert the light curve to the unabsorbed flux (0.3-10~keV) using the conversion factor of 3.7 $\times$ 10$^{-11}$ erg cm$^{-2}$ ct$^{-1}$.
A basic comparison with the closure relations \citep{Zhang2004} shows
that the measured spectral and temporal parameters 
could be consistent with a simple forward shock model, in either slow cooling or fast cooling regime, provided that the power-law index $p$ of the accelerated electron distribution is $p\approx$2.6.

\begin{figure}
\includegraphics[scale=0.17,trim=2 1 27 13, clip]{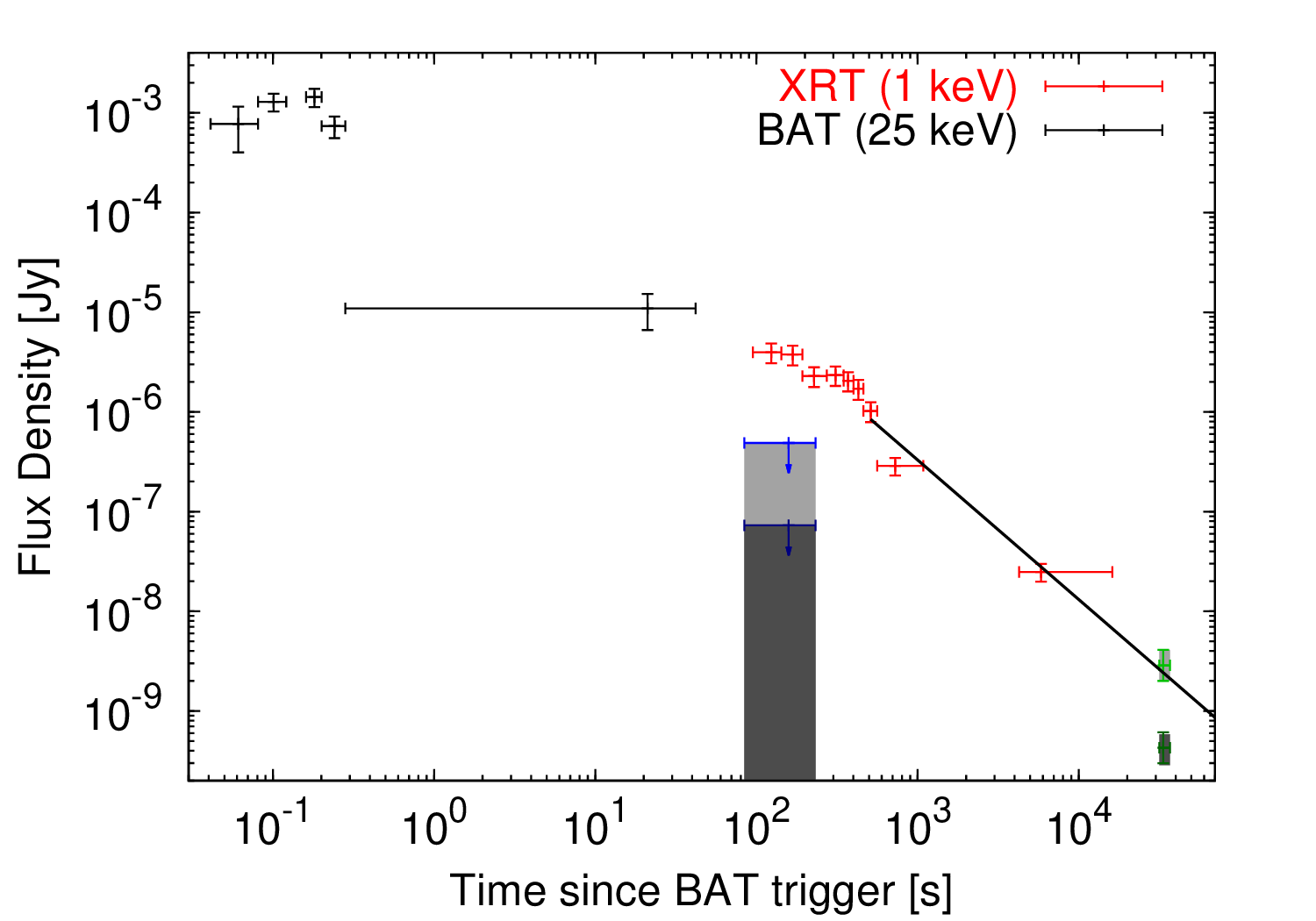}
\caption{Comparison between {\it Swift}/XRT data sample and the optical flux extrapolated to 1 keV. 
The solid line represents the best fit model for the X-ray light curve. 
The light blue and dark blue upper limit shows the UVOT early observations in the $white$ filter assuming the most extreme cooling break between X-ray and optical (light gray), and no cooling break 
(dark gray), respectively. The light and dark green points show the $i$-band  
Gemini observation at 0.38 days for the two spectral assumptions, respectively.
\label{xray-opt_extrapolation}}
\end{figure}

\begin{table*}
 	\centering
 	\caption{Short GRBs at high redshift.}
 	\label{tab:highz}
 	\resizebox{.95\textwidth}{!}{%
 	\begin{tabular}{lccccccccccc}
    \hline
    GRB name & $z^{a}$ & EE & Signif. & $T_{90}^{b}$ & HR$^{c}$ & $E_{\gamma,iso}^{d}$ & $f_{NC}^{e}$ & Pop. Age & log(M/M$_{\odot}$) & SFR & Offset \\
        &          &                   &  [SNR] & [s] &  & [erg] & \% & [Gyrs] &  & [M$_{\odot}$/yr] & [kpc] \\  
   \hline
051210$^{f}$ & $>$1.4 & yes & 3.3 & 1.30$\pm$0.30 & 1.9$\pm$0.5 & $> 2 \times 10^{50} $& $82^{+10}_{-60}$ & $0.15^{+0.13}_{-0.08}$ & $9.4^{+0.3}_{-0.4}$ & $17.0^{+9.0}_{-5.0}$ & $>17$\\
060121 & $\gtrsim$ 1.7 & yes & 3.3 & 1.97$\pm$0.06 & 1.5$\pm$0.2 & $\gtrsim$ 3.6 $\times 10^{52}$ & $17^{+14}_{-15}$ & - & - & - & $<1.4$\\
090426A & 2.609 & no & - & 1.24$\pm$0.25 & 1.1$\pm$0.3 & $1.6_{-0.2}^{+0.2} \times 10^{51}$ & $10^{+14}_{-6}$ & $0.08^{+0.05}_{-0.02}$ & $8.1^{+0.2}_{-0.1}$ & $4.3^{+2.0}_{-2.0}$ & $0.5 \pm 0.2$\\
111117A & 2.211 & no & - & 0.46$\pm$0.05 & 2.8$\pm$0.6 & $2.8_{-0.4}^{+0.4} \times 10^{50}$ & $96^{+3}_{-5}$ & $0.23^{+0.36}_{-0.16}$ & $9.6^{+0.3}_{-0.4}$ & $17.4^{+9.3}_{-6.6}$ & $8.5 \pm 1.7$ \\
120804A & $1.0^{+0.4}_{-0.2}$ & yes & 4.2 & 0.81$\pm$0.08 & 1.6$\pm$0.1 & $\approx 1.4 \times 10^{51}$ & $36^{+11}_{-19}$ & $0.33^{+0.13}_{-0.12}$ & $10.2^{+0.2}_{-0.2}$ & $40.0^{+33.0}_{-28.0}$ & $2.3 \pm 1.3$\\
121226A & $1.8^{+0.2}_{-0.1}$ & no & - & 1.01$\pm$0.20 & 1.4$\pm$0.4 & $\approx 6 \times 10^{50}$ & $28^{+10}_{-16}$ & $1.00^{+0.48}_{-0.44}$ & $10.2^{+0.1}_{-0.1}$ & $12.7^{+9.4}_{-7.6}$ & $<20$\\
160410A & 1.717 & yes & 6.9 & 96$\pm$50 & 2.3$\pm$0.5 & $1.2_{-0.1}^{+0.1}\times 10^{51}$ & $59^{+19}_{-22}$ & - & - & - & - \\
181123B & 1.754 & yes & 4.0 & 0.26$\pm$0.04 & 2.4$\pm$0.6 & $2.6^{-0.6}_{+0.6} \times 10^{50}$ & $98^{+2}_{-3}$ & $0.50^{+0.90}_{-0.30}$ & $10.0^{+0.2}_{-0.2}$ & $18.1^{+15.9}_{-9.3}$ & $5.1 \pm 1.4$ \\
    \hline
    \end{tabular}%
    }
\begin{flushleft}
\quad \footnotesize {$^a$ References: GRB 051210 \cite{Berger2007}, GRB 060121 \cite{deUgartePostigo2006}, GRB 090426 \cite{Levesque2010}, GRB 111117A \cite{Selsing2018}, GRB 120804A \cite{Berger2013}, GRB 121226A \cite{Selsing2016}, GRB 181123B \cite{Paterson2020}}\\
\quad \footnotesize {$^b$ $T_{90}$ values were retrieved from the \textit{Swift} BAT GRB catalog 
\citep{Lien2016}.}\\
\quad \footnotesize{$^c$ The hardness ratio is reported only for the main short duration peak, and computed as the ratio between the fluences in the 50-100 keV and 25-50 keV energy range. For GRB 060121, the value was retrieved from \cite{deUgartePostigo2006} and is the fluence ratio between the 100-300 keV and 50-100 keV energy range.}\\
\quad \footnotesize{$^d$ $E_{\gamma,iso}$ is derived in the 15-150 keV energy band (rest frame); except for the HETE-2 GRB 060121, for which it was derived in the 2-400 keV energy range (observer frame).}\\
\quad \footnotesize{$^e$ Probability that the GRB belongs to the class of non-collapsar events. Derived using the equations presented in \citet{Bromberg2013}. }\\
\quad \footnotesize{$^f$ The enhanced XRT position does not include the putative host galaxy proposed by \citet{Berger2007}, from which this redshift constraint is derived. The chance probability for this GRB/galaxy association is $\approx$15\%.}\\
\end{flushleft}
\end{table*}

\subsection{Search for extended emission in other high-\textit{z} sGRBs}

Fostered by the detection of EE in GRB181123B, 
we follow the same procedure to analyze the BAT data obtained for other sGRBs (see Table~\ref{tab:highz}) associated to high-$z$ ($z$\,$>$1) host galaxies.
These include GRB~051210 \citep{laparola05, Berger2007}, GRB060121 \citep{Donaghy2006, deUgartePostigo2006}, GRB~090426 \citep{Levesque2010, Antonelli2009}, GRB111117A \citep{Sakamoto2013, Selsing2018}, GRB120804A \citep{Berger2013} and GRB160410A \citep{Selsing2016}. 
To these known cases, we add GRB~121226A for which we derive $z$\,$\approx$1.8 based on the analysis of its host galaxy photometry (see Sect. 3.4). 
We do not include GRB~150423A because the reported redshift of 1.39 \citep{Malesani2015} is considered uncertain due to the low significance of the observed spectral features (D. Malesani, priv. comm.) and GRB~190627A at $z$\,$\sim$1.9, initially classified as a short burst, but later found to have a longer duration and a soft spectrum \citep[HR $\sim$ 0.7;][]{Lien2016}. 

Our analysis  yields a positive detection of EE in three cases: GRB~051210, GRB~160410A and GRB~120804A. In the latter case, we verified that the presence of a bright hard X-ray source (Sco X-1) within the BAT field of view does not significantly contaminate the source count rate. 
We do not confirm the tentative identification of EE in GRB121226A \citep{Pandey2019}, 
although we note that the definition of EE given in \citet{Pandey2019} differs from the one adopted in this work. 
As a sanity check we also tested our pipeline on two low-redshift bursts \citep[e.g. GRB 050509B and GRB 051221A; ][]{Gehrels2005,Burrows2006} and find no evidence of EE in the BAT data. 

For each high-$z$ burst we compute the $E_{\gamma,iso}$ (15-150 keV, rest frame) and the hardness ratio (HR), defined as the fluence ratio between the ranges 50-100 keV and 25-50 keV. We add to our sample GRB 060121, for which evidence of EE was observed in the High Energy Transient Explorer 2 (HETE-2) data \citep{Donaghy2006}. In this case, the HR reported in Table~\ref{tab:highz} refers to the ratio between the 100-300 keV and 50-100 keV fluences \citep{deUgartePostigo2006}. The results are summarized in Table~\ref{tab:highz}.\\

\section{Results} \label{sec:results}

\subsection{Evidence of extended emission in GRB181123B}
\label{sec:eemission}

In Section~\ref{sec:bat_analysis}, we find that a long-lasting signal, compatible with EE, is seen in the BAT image below 50 keV. Here we show that the tail of this EE is also visible in the early ($\lesssim$600 s) XRT light curve. 
We use the optical emission as a tracer of the standard forward shock afterglow, and examine whether the observed X-rays belong to the same emission component.

We consider the early UVOT upper limit  
\textit{Wh} $>$ 21 AB mag at 158.5~s \citep{Oates2018}
and the optical detection $i$=25.1 $\pm$ 0.4 at 0.38~d \citep{Paterson2020}.
After correcting for a Galactic extinction of $E(B-V)$=0.03~mag \citep{SCHAFLY2011} and the expected absorption due to the Lyman beak at
$z\approx$1.75, 
we extrapolate the optical fluxes to 1~keV assuming two different spectral shapes. In the first case, we assume a simple power-law spectrum connecting the X-ray and optical data with the spectral index defined by the X-ray spectra, $\beta_X$ = $\Gamma_X$ -1 $\approx$1.0 (Section~\ref{sec:xrt_analysis}).
In the second case, we assume a cooling break between the optical and X-rays with the optical index defined by the closure relations \citep[
$\beta_{o}$=$\beta_{X}$-0.5;][]{Zhang2004}. The highest X-ray flux is obtained assuming a cooling break at $\nu_{c}$=0.3 keV (right below the XRT energy band). 

Figure~\ref{xray-opt_extrapolation} shows the comparison between the observed X-ray light curve and the extrapolation of the optical data. 
To guide the eye we also show the power-law decay of the late X-ray afterglow (solid line in Figure~\ref{xray-opt_extrapolation}).
If the optical and X-ray emission are produced from the same forward shock, then the X-ray light curve should lie within the shaded regions of the optical data. 
This is observed for the \textit{i}-band data point at 0.4~d (the extrapolation lies exactly on the solid line when we assume a cooling break between X-ray and optical), but it is not seen at earlier times. 
Figure~\ref{xray-opt_extrapolation} shows that the extrapolation of the UVOT limit substantially underpredicts the observed X-ray emission at $\approx$150~s.
By assuming a Milky Way extinction law \citep{Fitzpatrick1999}, 
only a dust extinction as high as $A_V$\,$\approx$2 could reconcile the X-ray afterglow with the optical extrapolation. This value is highly unusual for short GRBs 
and it is not consistent with the limits on intrinsic absorption 
placed by the X-ray spectrum.
Therefore, although the X-ray dataset could be consistent with a simple forward shock model (Section~\ref{sec:xrt_analysis}), the joint analysis of the X-ray and optical data reveals an additional emission component in the X-ray band, not related to standard forward shock emission. We therefore suggest that the early X-ray light curve is consistent with the detection of continued high-energy emission in BAT, and likely shows the
final tail of the EE.

\begin{figure}
\includegraphics[height=4cm,width=4cm,trim=4 7 8 12, clip]{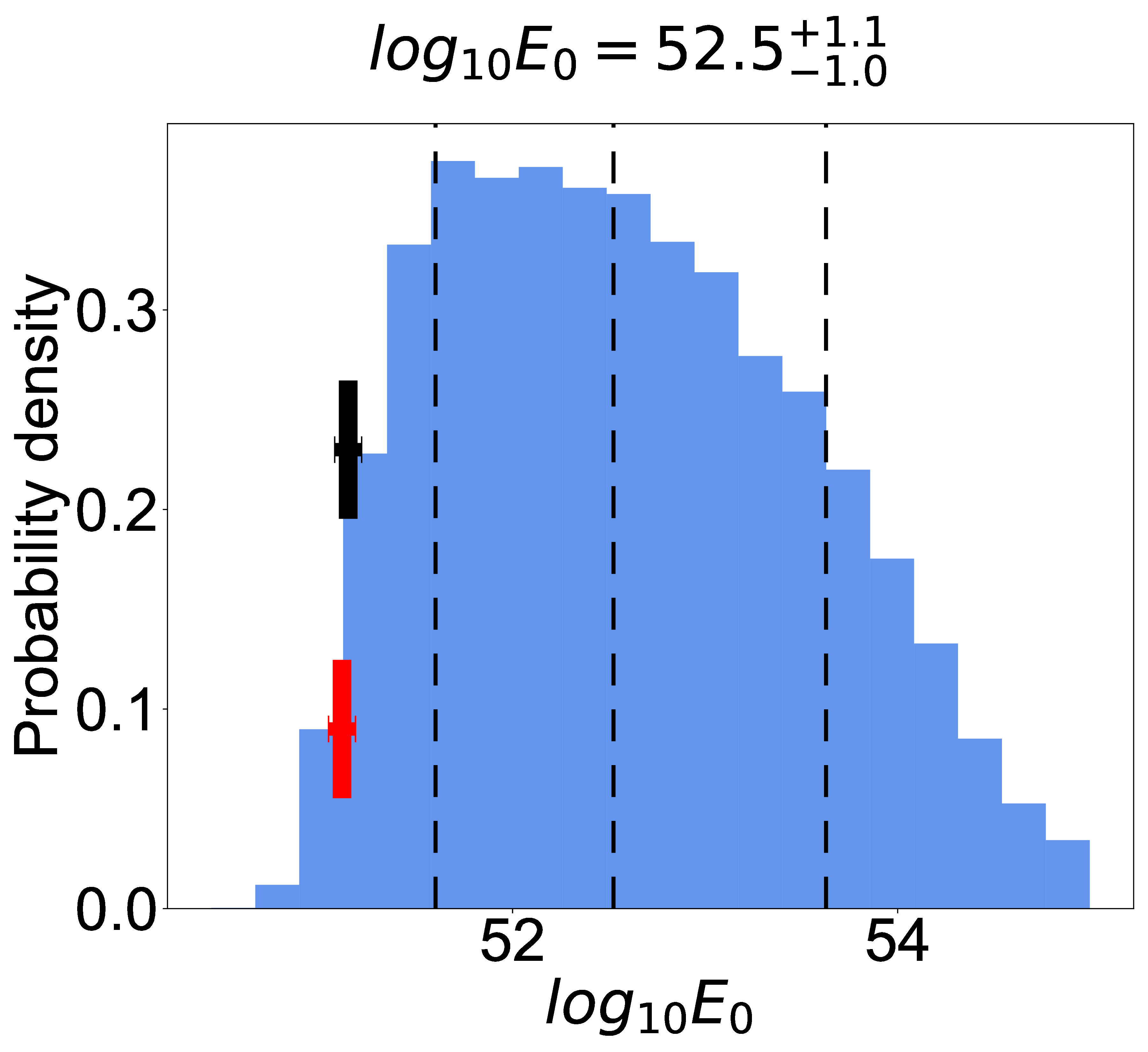}
\hspace{0.3cm}
\vspace{0.3cm}
\includegraphics[height=4cm,width=4cm,trim=4 7 8 12, clip]{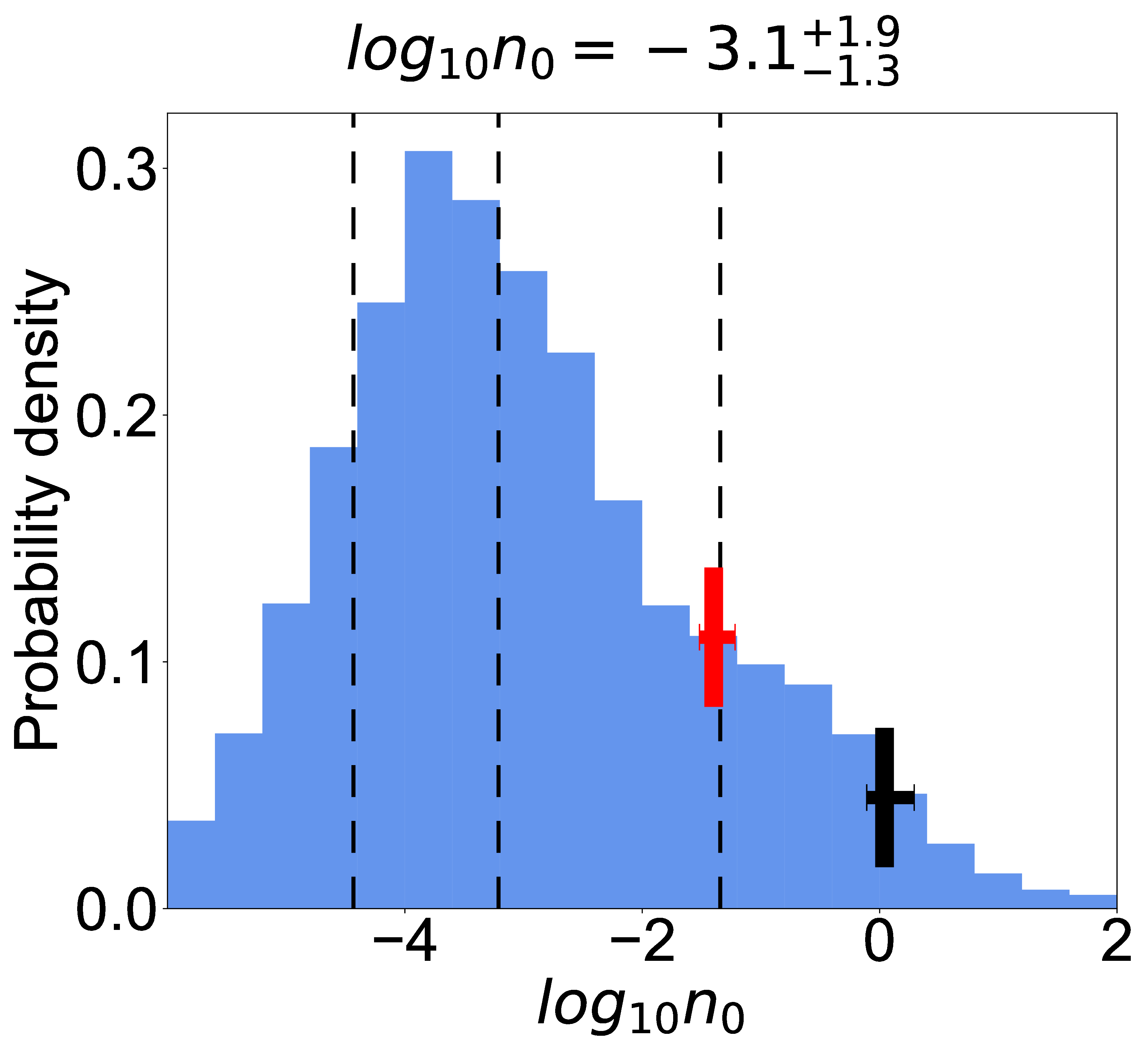}
\includegraphics[height=4cm,width=4cm,trim=4 7 8 12, clip]{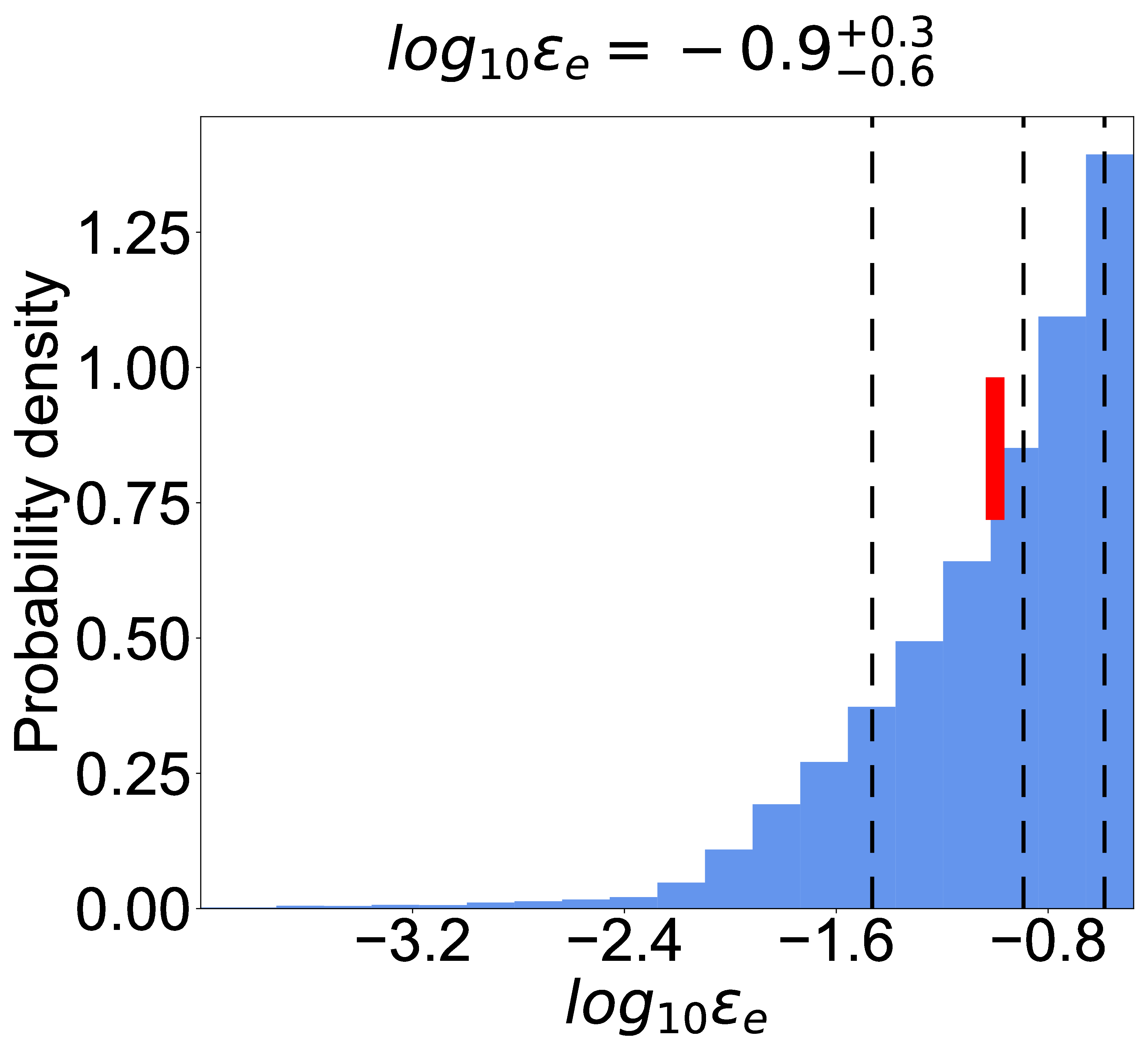}
\hspace{0.3cm}
\includegraphics[height=4cm,width=4cm,trim=4 7 8 12, clip]{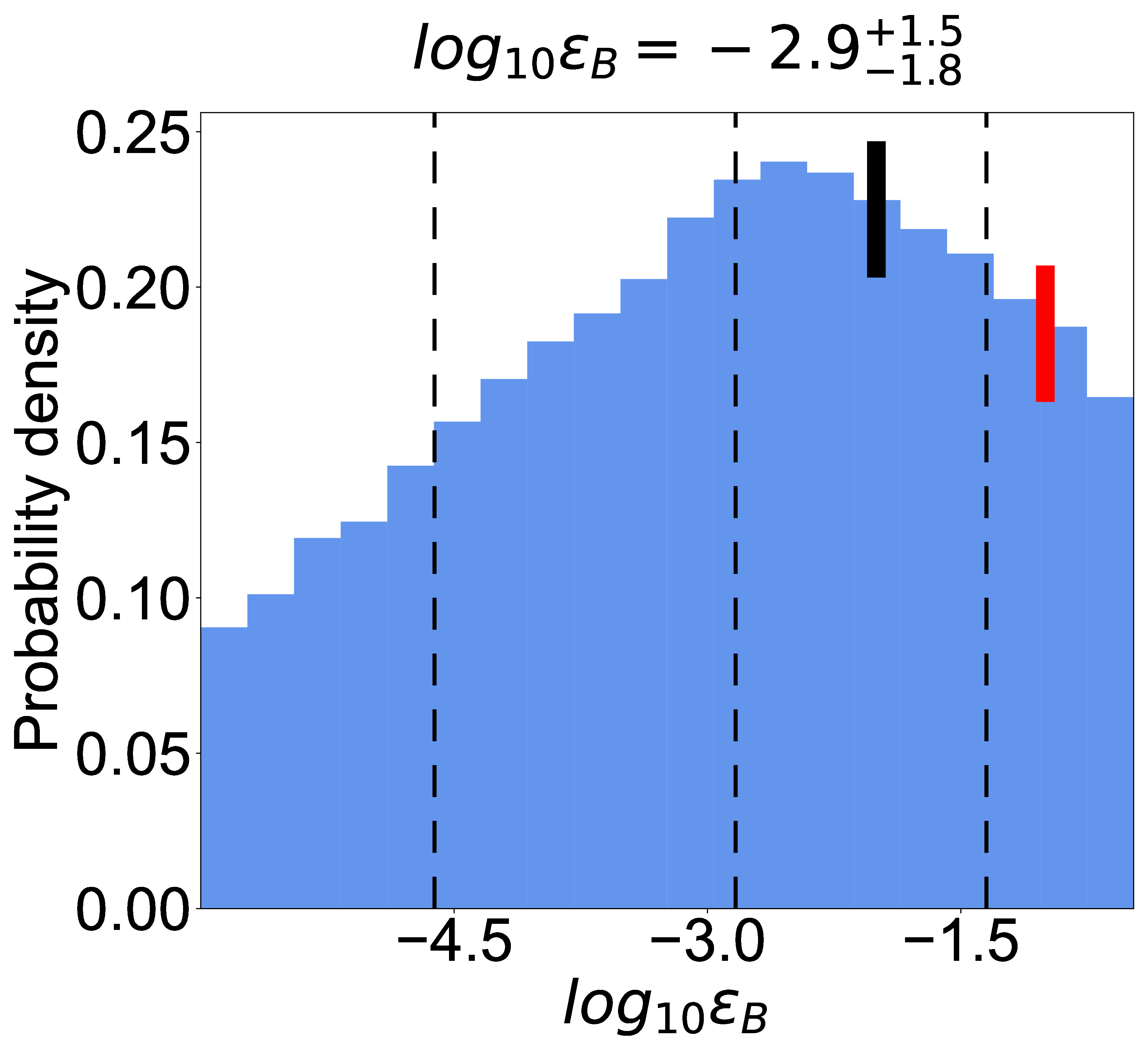}
\begin{center}
\includegraphics[height=4cm,width=4cm,trim=4 7 8 12, clip]{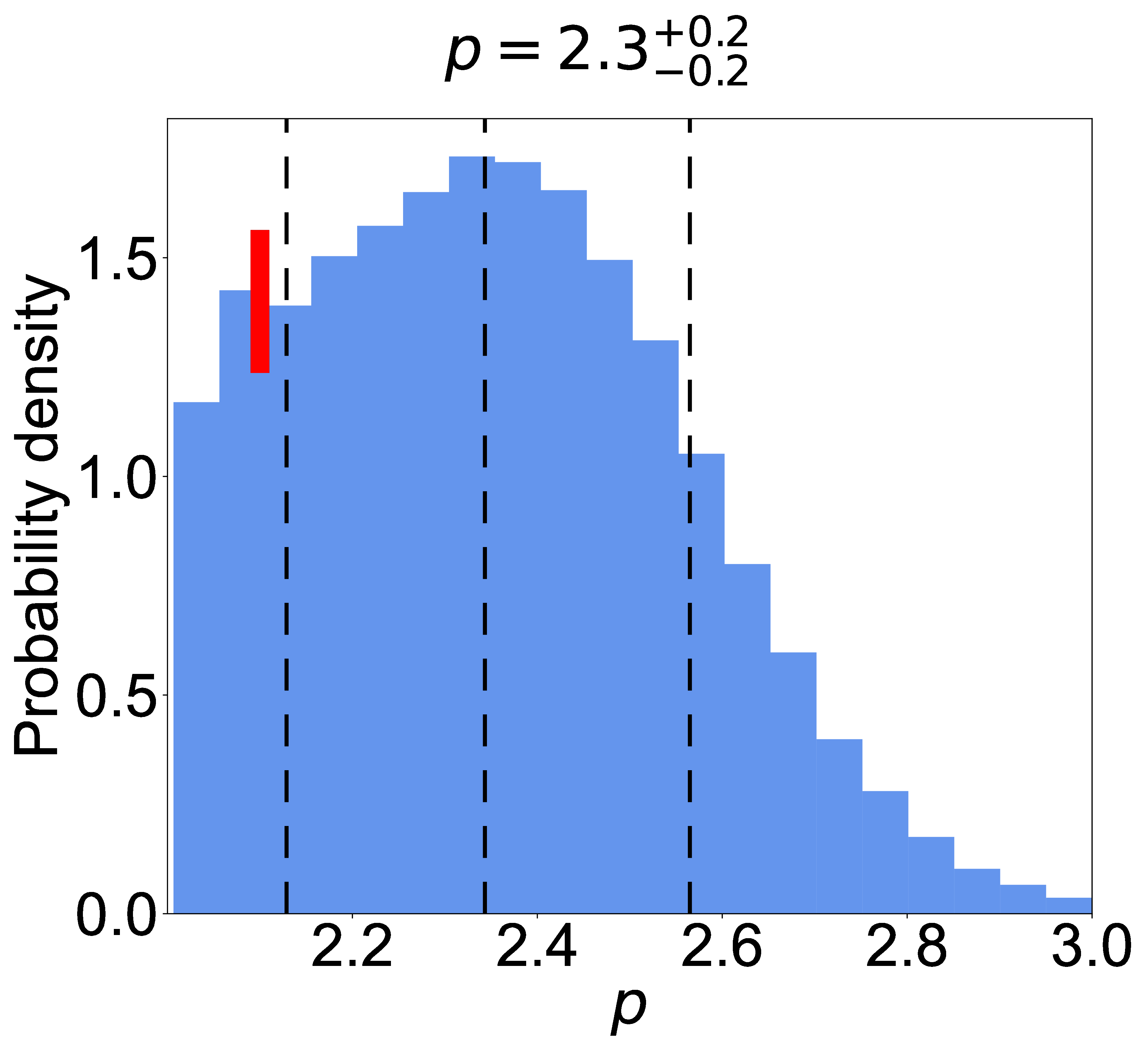}
\end{center}
\caption{Posterior distributions of the afterglow best fit parameters for GRB 181123B. Dashed lines show the median and the 1~$\sigma$ contours for each parameter
For comparison, we show the results of \cite{Paterson2020} for $\epsilon_{e}$=$\epsilon_{B}$=0.1 (red mark) and $\epsilon_{e}$=0.1, $\epsilon_{B}$=0.01 (black mark).
}
\label{afterglow-fit}
\end{figure}
 
\subsection{Afterglow constraints}\label{sec:afterglow}

In order to derive the afterglow parameters we remove the initial 600~s of X-ray data, which according to our analysis are contaminated by the extended emission. 
We include the late optical detection $i$\,$\approx$ 25.1 AB mag, additional constraints from the near-infrared ($J$ $>$23.2~AB mag), and the radio observations \citep{Anderson2018}.
Data from the Australian Telescope Compact Array (ATCA) were downloaded from the public archive and analyzed following \citet{Ricci2020}. Observations were taken in the 4cm band
in the 6B array configuration. At the GRB position, we derive upper limits of $\sim$45~$\mu$Jy at 5.5~GHz and 9~GHz at a median time of 17 hrs after the trigger.

We explore the parameter space using a standard fireball model and assuming a top-hat jet. We neglect the lateral spreading effects expected for a more complex jet structure, since they are likely to be not relevant over the time span of these observations ($<$1~d).  
We use \textsc{afterglowpy} \citep{Ryan2020} to compute theoretical afterglow emission and fit our data using MCMC posterior sampling with \textsc{EMCEE} \citep{Foreman-Mackey2013}.
We use 100 walkers, for 20000 steps
and standard priors for the fit parameters
\citep[e.g.][]{Troja2018,Troja2020,OConnor2021}. 
The posterior distributions of the afterglow parameters are shown in Figure~\ref{afterglow-fit} assuming synchrotron emission with no significant synchrotron self-Compton (SSC) cooling. As the fitted dataset is limited and does not include early X-rays, SSC cooling corrections are expected to be sub-dominant (e.g. \citealt{Jacovich2020}). We have verified this by performing an additional fit, allowing for SSC cooling with no Klein-Nishina (KN) corrections (Thomson SSC). The lightcurve corresponding to SSC cooling with KN corrections, necessarily lies in between Thomson SSC cooling and the opposite extreme, of pure synchrotron. The inferred parameters for the Thomson SSC fit are reported in the Appendix. 
$\eta_{\gamma}$ decreases as expected when SSC is accounted for \citep{BNP2016}. However, the decrease in this case is rather small compared to the synchrotron only case and overall the inferred parameters change very little between the two fits.

As expected, this limited dataset can only weakly constrain the parameter space.
The circumburst density $n$ favors low density solutions, typical of sGRBs \citep{OConnor2020}, and high densities ($n$\,$\gtrsim$10 cm$^{-3}$) can be excluded at the 90\% confidence level. 
Other parameters, such as the electron index $p$=2.3$\pm$0.2, and the blastwave isotropic-equivalent energy,
log($E_{0}$/erg)
= 52.5$^{+1.1}_{-0.9}$, 
are consistent with typical sGRB afterglows \citep{Fong2015}. In particular, the implied efficiency of prompt $\gamma$-rays relative to the kinetic energy is $\eta_{\gamma}$\,$\sim$\,10\%, consistent with values estimated in other GRBs \citep{Nava2014,Beniamini2015,BNP2016}.
The electron energy fraction $\varepsilon_e$ appears well determined as $\varepsilon_e$ $>$ 0.01, consistent with typical GRBs \citep{BvdH2017}, whereas the magnetic  energy fraction $\varepsilon_B$ remains loosely defined between $10^{-5}$ and $\approx$0.1 \citep[as typically observed for other GRBs,][]{Santana2014, Zhang2015}.
For comparison, the solutions proposed by \cite{Paterson2020} 
are marked by the vertical bars in Figure~\ref{afterglow-fit}. 
By fixing several parameters ($p$, $\varepsilon_e$, $\varepsilon_B$) to arbitrary values, these solutions
probe the low probability tail of our posterior distributions and imply an unusually high radiative efficiency ($\eta_{\gamma}>$80\%). A much broader range of values is instead consistent with the observations, as expected for a poorly sampled afterglow. 

\subsection{GRB classification}\label{sec:simulations}

\subsubsection{Duration}

We find that most sGRBs located at $z\gtrsim$1 display a weak, long-lasting high-energy emission (see Table~\ref{tab:highz}). However, in all but one case (GRB~160410A), this emission is not picked up by the standard pipeline and the measured $T_{90}$ is $<$2\,s.
This may be interpreted as a true physical distinction 
between these bursts and other sGRBEEs, whose $T_{90}$ is longer than 2~s. 
However, we show that instrumental selection effects are a strong determining factor.

To illustrate the role of instrumental effects in the calculation of $T_{90}$, 
we consider the case of GRB~071227 with a measured $T_{90}$ of $143 \pm 48$~s \citep{Lien2016}.
Despite its long duration, this burst is classified as a short GRB with EE
based on the morphology of its gamma-ray lightcurve (a short spike followed by a weaker temporally extended tail), its environment, and the lack of SN emission to deep limits \citep{DAvanzo2009}. According to the analysis of \citet{Norris2011}, 
the gamma-ray properties of GRB~071227 are representative of the general population of sGRBEEs. Below, we show that the phenomenological classification of this burst as a sGRBEE would change under different observing conditions. 

We use the observed GRB light curve as input to simulate \textit{Swift}/BAT 
observations for different observing conditions (e.g. different redshift, background level). 
Our code takes into account the proper instrument response matrices and trigger algorithms \citep[][Moss et al., in prep.]{Lien2014},
and simulates light curves at various redshifts by calculating the distance and time-dilation corrections for the input light curve. The standard Bayesian blocks tool \citep{Scargle2013} is then ran on the simulated lightcurve to derive its $T_{90}$. 

\begin{figure}[ht!]
\includegraphics[scale=0.58]{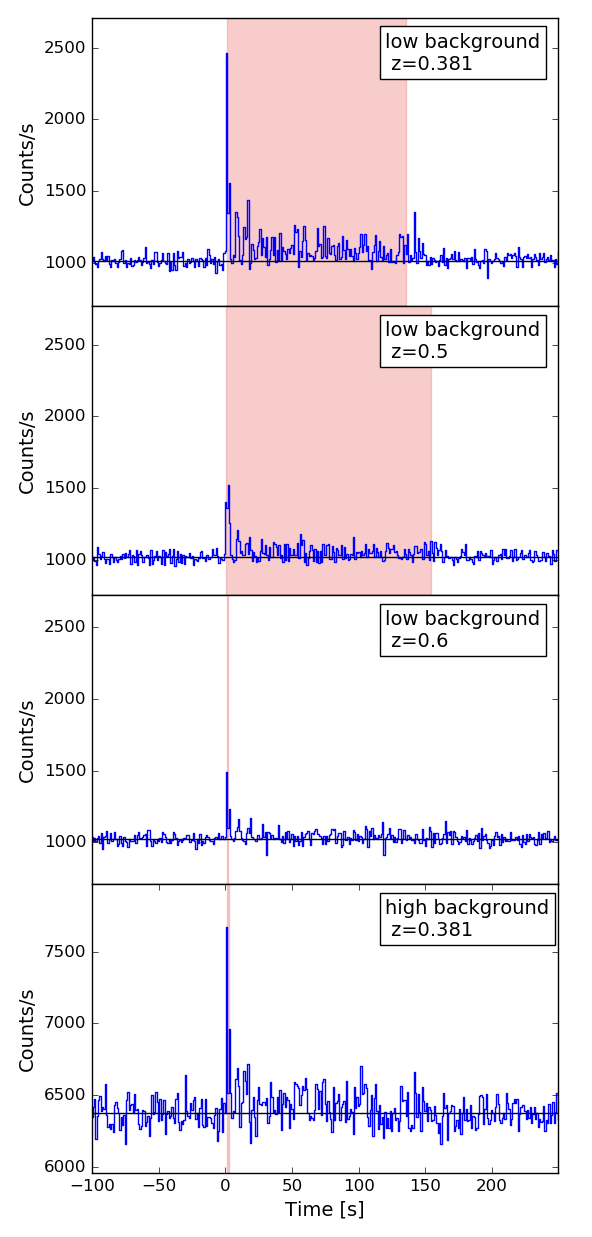}
\caption{Simulated light curves of GRB 071227, derived assuming different redshifts and average background levels. The estimated $T_{90}$ is shown by the shadow pink area and it varies from $T_{90} = 136$ to $T_{90} < 2 s$ moving from the observed redshift $z=0.381$ to $z=0.6$. The extended emission can not be detected in case of high background level.
}
\label{simulations}
\end{figure}

The results are shown in Figure~\ref{simulations}. 
We show GRB~071227 at three different redshifts:
its true value $z=0.381$, and two higher values $z=0.5$ and $z=0.6$. All these simulations were carried out assuming 
a low average background of $\sim1,000$ cts/s. 
The shaded area shows the $T_{90}$ interval, which recovers the presence of a temporally extended emission up to $z$=0.5. However, for $z$\,$\gtrsim$0.6 the EE becomes undetectable and the burst would be classified as a standard sGRB with $T_{90}<$2~s. 
Instead, the tail of EE would remain visible in the early X-ray light curve of GRB~071227, as also seen for GRB~181123B. Similarly to GRB~181123B, the other three high-$z$ sGRB with evidence for a weak EE (GRB 051210, GRB 160410 and GRB 120804A) show a phase of rapid decay in their early X-ray afterglow \citep[e.g.][]{laparola05}
consistent with the tail of the extended emission.
A joint analysis of the BAT+XRT data could therefore be more effective in recovering its presence \citep[e.g. ][]{Kisaka2017}.

In the bottom panel of Figure~\ref{simulations}, we show the effect of background variations on the detectability of EE. 
By using a typical background value of $\sim 6,000$ cts/s,  we find that the EE of GRB~071227 may be lost even for $z$=0.381. 
The same effect applies to observations with a different number
of active BAT detectors. This number has been steadily decreasing with time (cf. Figure 3 of \citealt{Lien2014}) and therefore the EE was more likely to be identified in the early years of the {\it Swift} mission. For example, 
over 24,500 detectors were active during the observations of GRB~071227, but only half of them ($\approx$12,200)
were on during the observations of GRB~181123B.

\subsubsection{Empirical correlations}

In addition to the duration and spectral hardness, other observables may 
aid in the GRB classification. 
For example, most long GRBs display a correlation between their isotropic gamma-ray energy release ($E_{\gamma,iso}$) and their rest-frame spectral peak energy ($E_{peak}$), known as the ``Amati'' relation \citep{Amati2002}. 
\begin{figure}[t!]
\begin{center}
\includegraphics[width=0.94\columnwidth]{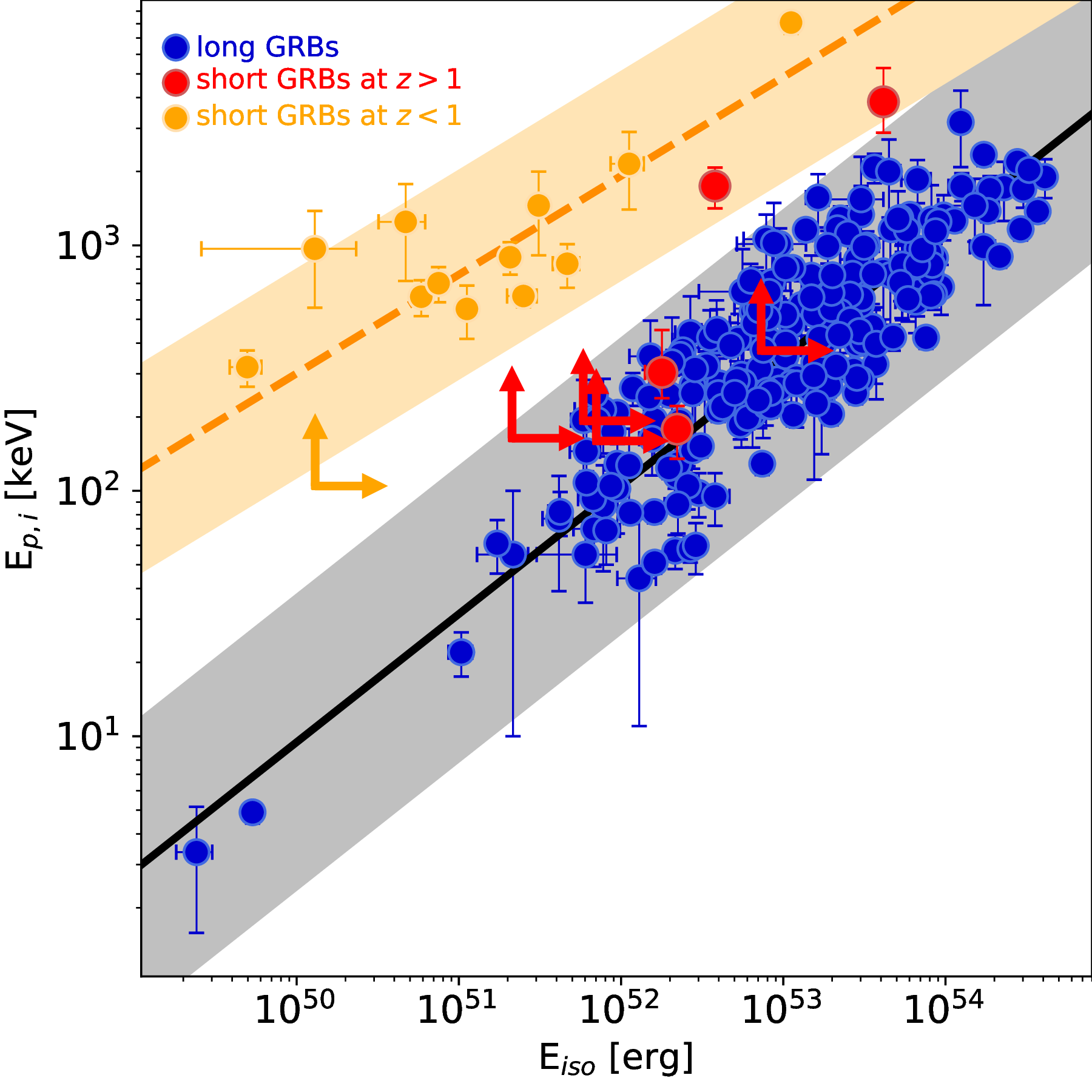}
\end{center}
\caption{GRBs in the $E_{peak}$-$E_{iso}$ plane including long bursts (blue), short bursts at $z<$1 (orange) and the subsample of sGRB at $z>$1 (red). Lower limits at the 90\% confidence level are shown by the arrows.
The solid (black) and dashed (orange) line shows the best fit obtained using the sample of long and short GRBs, respectively. The shadowed areas show the 3~$\sigma$ scatter for the 2 correlations. Figure updated from \cite{Amati2019}.
}
\label{epeiso}
\end{figure}
We verify whether the bursts of our sample follow the Amati relation or not, like most sGRBs. 
The spectral peak energies were retrieved from the literature \citep{Donaghy2006,GCN13614,GCN19288} and from the {\it Fermi}/GBM catalog \citep{vonKienlin2020} when available. Otherwise, only lower limits on $E_{peak}$ were derived from the analysis of the {\it Swift} BAT spectra.
We find that only GRB~111117A lies outside of the 3~$\sigma$ upper boundary (Figure~\ref{epeiso}), whereas 
GRB~090426A, GRB~120804A and GRB~160410A are consistent with the relation within 3~$\sigma$. GRB~051210, GRB~060121, GRB~181123B and GRB~121226A also fall within the correlation, but in these cases only lower limits on the peak energy and energy release are available. 

A correlation between the spectral lag and the peak luminosity is also observed for many long GRBs \citep{Norris2000}, and can be used to discriminate between the different GRB classes \citep[e.g.][]{Becerra2019}. For the GRBs in our sample, even when the lag is small,the large uncertainty in the lag measurement combined with the high luminosity of the gamma-ray emission makes these bursts consistent (within 3 $\sigma$) with the lag-luminosity relation observed for long bursts. 
The only exception is again GRB 111117A for which the lag is particularly well constrained (6.0 $\pm$ 2.4 ms) and places this burst
in the region populated by sGRBs.

\subsection{Environment}
\label{sec:hosts_study}

In order to further characterize the nature of these high redshift short GRBs and their EE,
we investigate their environments and compare them to the populations of long and short GRBs. 
Photometric observations of their host galaxies were retrieved from the literature \citep{LeiblerBerger2010, Levesque2010, Sakamoto2013, Selsing2018, Berger2014, Paterson2020}
and homogeneously modelled with \textsc{Prospector} \citep{Johnson2017} using the same methodology described in \cite{OConnor2021}.
In this work, we also present new photometric measurements for the host galaxy of GRB~121226A observed with the 8.1m Gemini South Telescope in the $gZY$ filters 
and the Very Large Telescope in the $J$ and $K$ filters. Magnitudes, calibrated to nearby PanSTARRS \citep{Chambers2016} and 2MASS \citep{Skrutskie2006} sources, are reported in Table~\ref{tab:121226A}. To model the spectral energy distribution of this specific GRB we also used the $r$ and $i$ band  photometry reported by \cite{Pandey2019}. 

\begin{table}[t!]
 	\centering
 	\caption{GRB~121226A Host Galaxy Photometry.}
 	\label{tab:121226A}
 	\begin{tabular}{lcc}
 	\hline
Filter & Telescope/Instrument & AB mag$^{a}$ \\
\hline
{\it g} & Gemini/GMOS-S & 24.3 $\pm$ 0.4 \\
{\it Z} & Gemini/GMOS-S & 23.99 $\pm$ 0.12 \\
{\it Y} & Gemini/GMOS-S & 23.62 $\pm$ 0.23 \\
{\it J} & VLT/HAWK-I & 22.97 $\pm$ 0.10 \\
{\it K} & VLT/HAWK-I & 22.43 $\pm$ 0.09 \\
\hline
    \end{tabular}%
\begin{flushleft}
\quad \footnotesize{$^{a}$ These values have been corrected for Galactic extinction due to the reddening of $E(B-V)$=0.05 \citep{SCHAFLY2011}.}\\
\end{flushleft}
\end{table}

\begin{figure}[b!]
\begin{center}
\includegraphics[width=0.95\columnwidth]{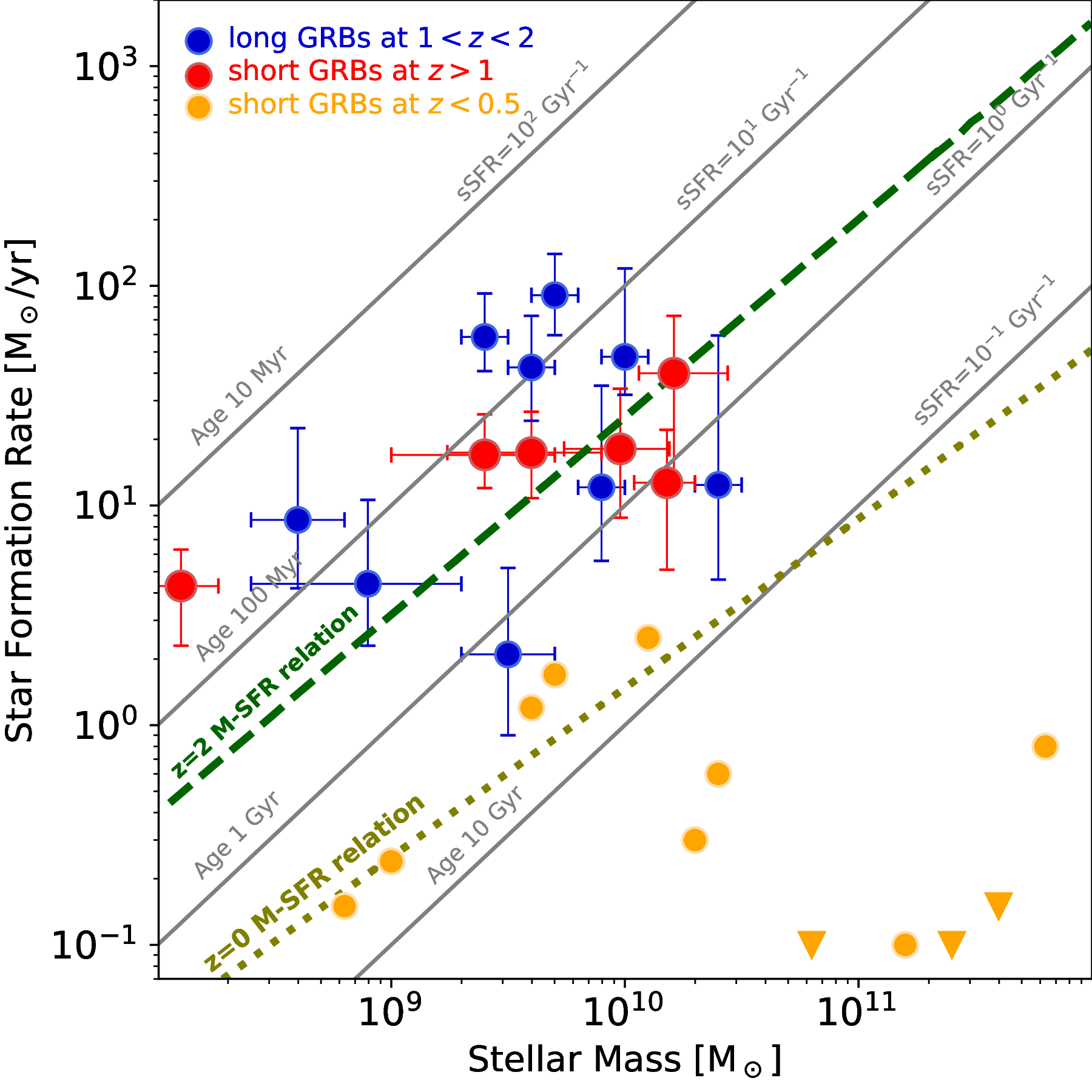}
\end{center}
\caption{In this plot we show the SFR and stellar mass obtained for the subsample of high-$z$ short GRBs as described in Section~\ref{sec:discussion} (red points in the figure). Blue points represent the values associated with the host galaxies of a sample of long GRBs detected between $z$=1 and $z$=2, retrieved from \cite{Palmerio2019}. Dashed green lines show the SFR-Mass relation for galaxies at $z\sim$2 \citep{Daddi2007} and $z\sim$0 \citep{Elbaz2007}. 
For comparison we also include in orange the values related to short GRBs host galaxies at
$z<$ 0.5 
\citep[from ][]{Berger2014}.\\
}
\label{galaxy_properties}
\end{figure}

We input the redshift as a fixed value if measured through spectroscopy, and leave it as a free parameter otherwise (GRB~051210, GRB~120804A and GRB~121226A). Two events (GRB~060121, GRB~160410) were not included in the analysis due to the limited dataset available. 
The results are listed in Table~\ref{tab:highz} and displayed in Figure~\ref{galaxy_properties}. 
We compare them with the sample of long GRBs in a similar redshift range \citep{Palmerio2019}
and find significant overlap in both stellar mass and star formation rate (SFR). 
Both groups of bursts are consistent with the $SFR$-$M$ obtained for GOODS catalog galaxies at $z$=2 
\citep{Daddi2007}. 
The orange points in Figure~\ref{galaxy_properties} show the distribution of sGRB with $z$\,$<$0.5 \citep[retrieved from ][]{Berger2014}. Some of these bursts are consistent with the $SFR$-$M$ relation at $z$\,$=$0 (dot-dashed line), whereas a large fraction resides in galaxies with larger stellar masses and lower SFRs. 

In order to compare the two populations of sGRBs at low and high redshifts, we correct for the evolution of star formation across cosmic time, as described in \citet{Bochenek2020}.  
For each galaxy, we rescale the SFR and mass so that they lie at the same distance (in units of standard deviations) from the $SFR$-$M$ relation at $z$=2 for $M$=$M_{z}$ (the host measured mass) and at $z$=0 for $M$=$M_{0}$ (the mass associated with the same quantile for the distribution at $z\sim$0). 
These scaled values are then compared with sGRB galaxies at $z$\,$<$0.5 using the two dimensional Peacock's test \citep{Peacock1983} which compares the mass and SFR distributions of the two samples, obtaining a p-value of 0.11. 
A much stronger similarity (p-value$\sim$0.7) is observed with the hosts of long GRBs at 1.0$<z<$2.0.

In Table~\ref{tab:highz} we report also the projected offsets for our sample of GRBs with respect to the host galaxy, updated from \citet{Fong2013}. Only GRB~051210, for which the host galaxy association is highly uncertain,  is particularly offset from its candidate host. All the other bursts have values consistent with the typical offset distribution of long GRBs which extends from $\sim$0.1 to $\sim$10 kpc with a median value around 1 kpc \citep{Lyman2017}. 

A useful tool to distinguish between the different classes could be the host metallicity. 
For instance,  sGRBs hosts track the metallicity distribution of field galaxies \citep{Berger2014}, while \cite{Palmerio2019} find that long GRBs at z$>$1 tend to avoid regions of high metallicity. Further studies on the metallicity of our sample of GRBs could clarify their classification.
\section{Discussion} \label{sec:discussion}
 
The results of the analysis presented in this paper suggest that a temporally extended emission can be identified in a significant fraction (60\%) of sGRBs at $z$\,$\gtrsim$1. This is much larger than the equivalent fraction that has been estimated in the general sGRB population \citep[$\lesssim$25\%, ][]{Norris2010}.
We suggest three main interpretations of this result and discuss some of their implications.

\subsection{Long GRB Impostors}

The first possibility is the `impostor' scenario. Namely, that a large fraction of sGRBs at $z>1$ are not in fact the result of compact binary mergers. They may be collapsars or potentially even a different population altogether. 
Under this interpretation, the apparent EE like component is not equivalent to the EE seen in lower redshift sGRBs and may, for example, be the regular prompt emission of a long GRB that after $\sim 1$\, s has become softer and more difficult to detect with BAT.
This interpretation is supported by spectral studies of GRBs, which find that when comparing the first 1-2 seconds of long GRBs, they are consistent with being drawn from the sGRB population \citep{Ghirlanda2009}, while at later times, long GRBs are significantly softer. Negligible spectral lags are also measured in some long GRBs \citep{Norris2011}, including those securely associated to supernovae \citep[e.g. GRB~091127,][]{Troja2012}.  
This interpretation is also consistent with the observed properties of the host galaxies (Figure~\ref{galaxy_properties}), showing no clear difference between long GRBs and sGRBs at $z$\,$>$1.

If the impostor scenario is correct, the true number of sGRBs at redshifts $z>1$ could be smaller than previously estimated. This would favor a shallower DTD, with a much larger fraction of systems with long delay times. This is no trivial requirement, given that there are $\gtrsim 30$ sGRBs with a measured $z<1$. 

In our analysis we find three sGRBs with $z$\,$>$\,1 that show no EE, GRB~121226A, GRB~111117A and GRB~090426A. Based on the observed gamma-ray emission, only GRB~111117A has a high probability of being a compact binary merger (see Table~\ref{tab:highz}).
GRB~121226A is not well constrained, whereas the nature of GRB~090426A is rather ambiguous and more likely associated to a massive star progenitor \citep{Levesque2010}. 
This leaves us with only one bona-fide sGRB at $z$\,$>$1 out of the 8 identified so far. 
To discuss the implications of this scenario, we adopt the most conservative assumption that is consistent with this ratio (i.e. one that maximized the number of sGRBs at $z>1$). Namely, we assume
that the large sample of sGRBs without redshift ($\approx$100 events) is composed by high-$z$ events and that the same ratio (1 sGRB, 7 long GRB impostors) applies to the entire observed population, we expect $\approx$15 sGRBs at $z>$1 within the {\it Swift} sample. 
Altogether this corresponds to $\approx$35-40\% of sGRBs at $z>1$. If modeled with a narrow width log-normal delay time distribution between star-formation and binary merger, this leads to very long delay times of order $3-3.5$\,Gyr  \citep{WP15}.
Taken at face value, this seems to be at odds with the observed population of binary neutron stars in the Galaxy, for which at least 40\% of the systems must have been born with delay times of $\lesssim 1$\,Gyr \citep{BP19}. Indeed out of the eight observed Galactic binary neutron stars that will merge within a Hubble time or less, not even one has a merger time as long as this. 
Such a long delay time would also be inconsistent with requirements from $r$-process evolution in the Milky Way at $\mbox{[Fe/H]}>-1$ \citep{Hotokezaka2018}, with the observation of $r$-process enriched stars in ultra faint dwarf galaxies \cite{Beniamini2016,Beniamini2016a} and with the large scatter of $r$-process abundances in extremely metal poor stars in the halo of the Milky Way \citep{Argast2004,Tsujimoto2014,Vangioni2016}. We refer the reader to \cite{BP19} to a more in depth discussion of these points.

However, we stress that this apparent discrepancy regarding the DTD may be at least in part due to a selection bias, as the detection of a sGRB and an assignment of redshift to a detected sGRBs becomes significantly more difficult at $z\sim 1$ relative to lower redshifts. This naturally leads to an artificial skewing of the observed sGRB redshift distribution to lower redshifts. Approximately, the bias should be such that it is $\sim$\,70 times less likely to detect and assign a redshift to a $z>1$ sGRB than to an sGRB at lower redshift in order to regain consistency between the DTD inferred from $r$-process / Galactic binary neutron stars and that inferred from sGRBs, assuming the majority of the EE sGRBs at $z>1$ are `impostors'. Fully modeling this selection bias is an involved work onto its own which would require knowledge of (a) the luminosity function (which is not directly determined from observations, as it is convoluted with the rate and delay time distribution), (b) the distributions of spectral parameters in the comoving frame (this again involves selection effects, when trying to infer based on the observed population) and (c) the recovery fraction (fraction of detected sGRBs for which $z$ will be determined). While the first two factors have been considered in previous studies \citep{WP15,Ghirlanda2016}, the latter has not yet been systematically studied.
A more complete understanding of the redshift determination bias will be crucial for determining whether there is any inconsistency between the impostor scenario and independent measures of the delay time distribution.

\subsection{Selection bias}

A variant of the selection bias above, and perhaps the simplest explanation, could be that these are regular sGRBEEs,
analogous to the population identified by \citet{Norris2006}, and they are more likely to be localized and assigned redshift than other sGRBs at similar distances.
Indeed, sGRBEEs tend to be brighter than canonical sGRBs \citep{Gompertz2020,Norris2011,Troja2008}, which increases the chance of an accurate localization, and hence a galaxy association. If this interpretation is correct, the high fraction of sGRBEEs at $z>1$ 
is simply a selection bias. Selection effects could also be the culprit of the distributions of host galaxy parameters, as star-forming galaxies are more likely to be spectroscopically identified and small offsets favor the host galaxy identification.
Under this interpretation, there could be many sGRBs originating from $z>1$ and there is no discrepancy with the delay time distributions inferred from Galactic BNSs or $r$-process abundances. 
For instance, by assuming the same ratio of $\approx$25\% between sGRBs and sGRBEEs \citep{Norris2010} and a $\approx$50\% fraction of events residing in galaxies with low star formation, the number of bursts at $z$\,$>$1 would quickly rise to over 60, much larger than those at $z<$1.

\subsection{Redshift evolution}
The third possibility for explaining the high occurrence rate of sGRBEEs at $z$\,$>$1 is that this indicates an evolution of sGRBs with redshift.\footnote{\cite{Anand2018} found no difference in the redshift distribution of sGRBs with and without EE, however they studied an older sample of sGRBs with EE, not including the $z>1$ EE discussed in this work.}
Understanding the nature of such an evolution depends on the, yet to be resolved, underlying mechanism powering EE in sGRBs. 

Various interpretations have been suggested in the literature. One intriguing possibility is that sGRBEEs arise 
from NS-BH mergers, and the EE is the result of $r$-process heating on fall-back accretion \citep{Rosswog2007}.
\cite{Desai2019} have studied this model using relativistic simulations of mergers. Their findings suggest that due to their larger merged BH masses, and smaller ejecta electron fractions, NS-BH mergers are more likely than NS-NS mergers to lead to an $r$-process powered fallback driven EE phase. 
The higher incidence of sGRBEEs at $z>1$ may suggest that the fraction of sGRBs from NS-BH is significantly increased compared to lower redshifts.
However, this interpretation, much like the impostor scenario mentioned above, increases the tension with the DTD of NS-NS mergers, by suggesting that less binary NSs lead to a detectable sGRB at $z>1$. Furthermore, population synthesis studies suggest that the DTD of NS-BH should favor longer delays compared to NS-NS mergers \citep{Mapelli19}, which would imply a smaller fraction of NS-BH at $z>1$ rather than vice versa.

Other studies have suggested that sGRBEEs may be the result of magnetar central engines \citep{Bucciantini2012,Gompertz2013,Sarin2020}. At the very least, this requires the product of the NS-NS merger to produce a long lived (or indefinitely stable) NS. Under this interpretation, the large ratio of EE sGRBs at $z>1$ would suggest that such merger products are more common at high $z$. Estimating the cosmological evolution of BNS binary mergers is a complex task, involving various uncertain components even at $z=0$, such as common envelope evolution. Nonetheless, we note that at high redshift, the metallicity of progenitor stars is lower, leading to less mass loss during stellar evolution and eventually to overall heavier NSs, and merger products (that are more likely to quickly collapse to a BH). This will cause an opposite trend to the required one, namely that magnetars should be a less common merger product at high $z$.

\section{Conclusion}\label{sec:conclusions}

{\it Swift} BAT observations of the short GRB 181123B, located at $z$=1.754, reveal a faint temporally extended emission following the first short peak. The tail of this high-energy signal likely dominates the early ($<$600~s) X-ray afterglow. 
By re-analyzing the BAT data for a sample of sGRBs at $z$\,$>$1, 
we identify evidence for a weak extended emission in most events
(5 out of 8 in total).
We show that the detection of this extended signal depends on a combination of factors that include the instrument sensitivity and the source distance. 
Although GRB environment studies are important to distinguish between the two classes of bursts, the small number of events and observational biases make the classification harder at high redshifts. Indeed
the study of the host galaxy properties shows similarities in offset distribution, mass and star formation with the environment of long GRBs detected at similar redshifts. Studies of the host metallicities could offer a clearer discriminant tool, but these measurements are either missing or not well constrained for the GRBs in our sample.

We consider three main hypotheses to interpret these results. 
First, these bursts with extended emission could be misclassified long duration bursts. However, this would cause a discrepancy between the DTD inferred from our study and the one derived from Galactic BNS systems and $r$-process elements abundances. 
Second, these GRBs belong to a different population of bursts and the predominance of extended emission at high-$z$ could indicate a redshift evolution of their progenitors. However, this is not expected by most models. 
Last, the large fraction of sGRBs with extended emission could be the result of selection effects, making this population of sGRBs easier to detect and localize at higher redshifts.
A thorough investigation of the selection bias affecting the redshift measurements of sGRBs and future investigations oriented to study the progenitors of sGRBEEs will be important to reconcile the different estimates of the DTD and determine the contribution of BNS mergers to the cosmic $r$-process enrichment.

Given the uncertainty in the classification of these bursts, the third generation of gravitational wave detectors may play a crucial role in the secure identification of NS-NS and NS-BH mergers at high redshifts. 
Future gravitational wave detectors like the Cosmic Explorer \citep{Reitze2019} or the Einstein Telescope \citep{Punturo2010}
are expected to identify BNS mergers out to redshift $z$\,$\sim$2-3 with a rate of $\approx$10$^5$ events per year, 
and about 10\% of the triggers from mergers at $z \gtrsim$1.3 \citep{Maggiore2020}.

\section*{Acknowledgements}
This work was supported in part by the National Aeronautics and Space Administration through grant 80NSSC18K0429 issued through the Astrophysics Data Analysis Program. 
The research of PB was funded by the Gordon and Betty Moore Foundation through Grant GBMF5076.
Based on observations obtained at the international Gemini Observatory (PI: Troja), a program of NOIRLab, which is managed by the Association of Universities for Research in Astronomy (AURA) under a cooperative agreement with the National Science Foundation on behalf of the Gemini Observatory partnership.
This publication made use of data products supplied by the UK Swift Science Data Centre at the University of Leicester.

\clearpage 
\appendix

\section{GRB 181123B Afterglow fit including Synchrotron Self-Compton energy losses} \label{sec:appendix}

\begin{figure}[!h]
\includegraphics[height=4cm,width=4cm,trim=4 7 8 12, clip]{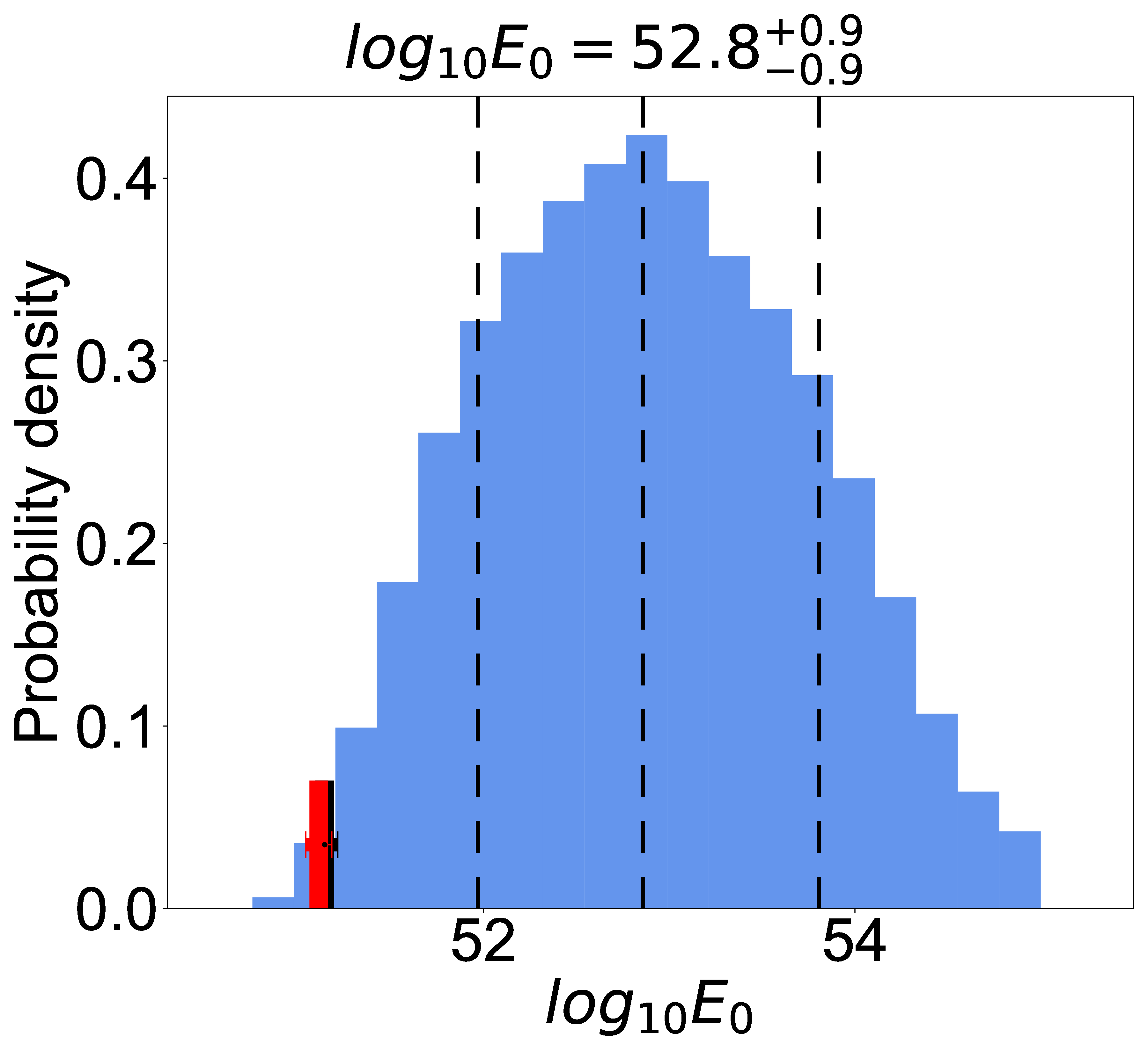}
\hspace{0.3cm}
\vspace{0.3cm}
\includegraphics[height=4cm,width=4cm,trim=4 7 8 12, clip]{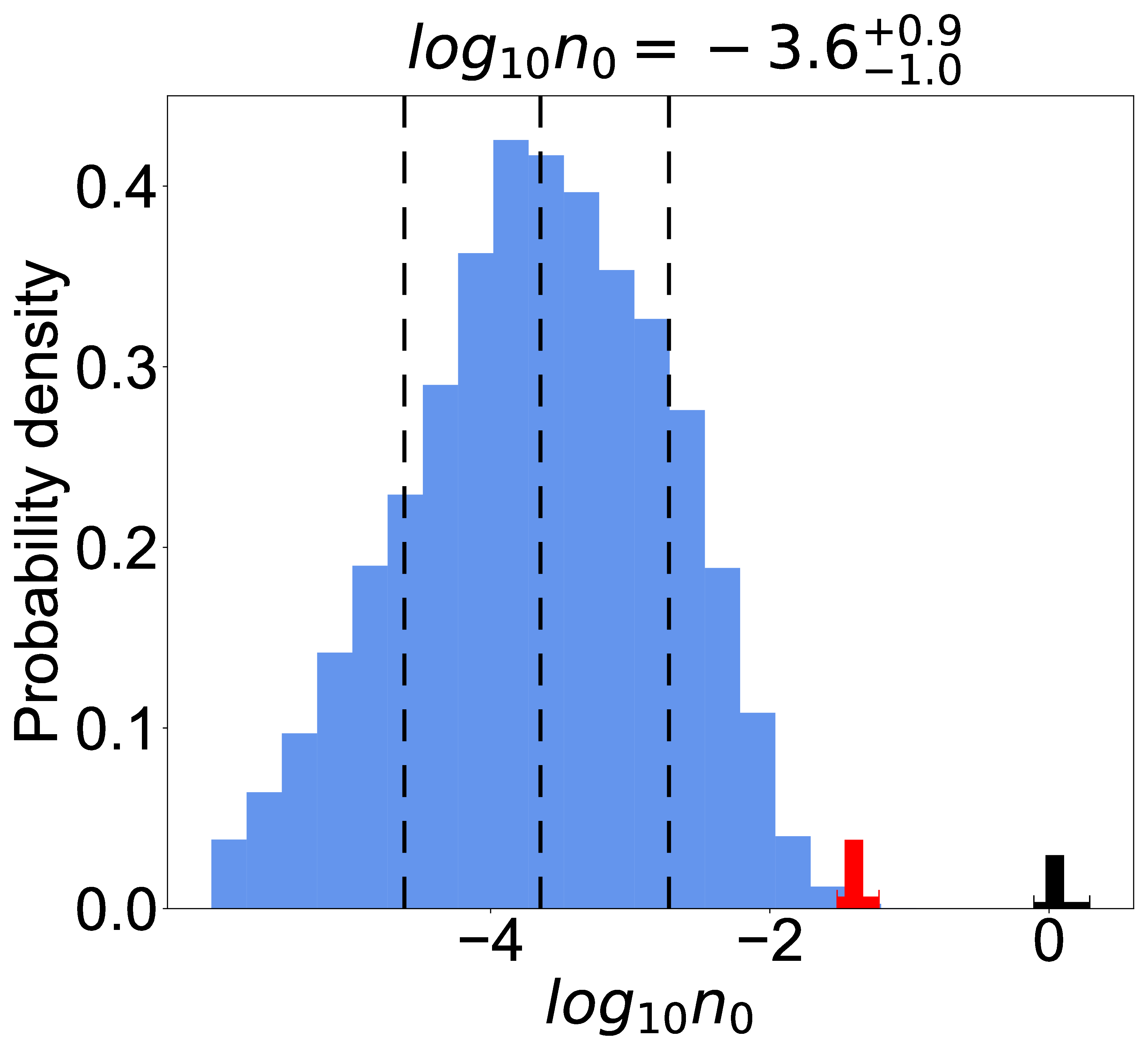}
\includegraphics[height=4cm,width=4cm,trim=4 7 8 12, clip]{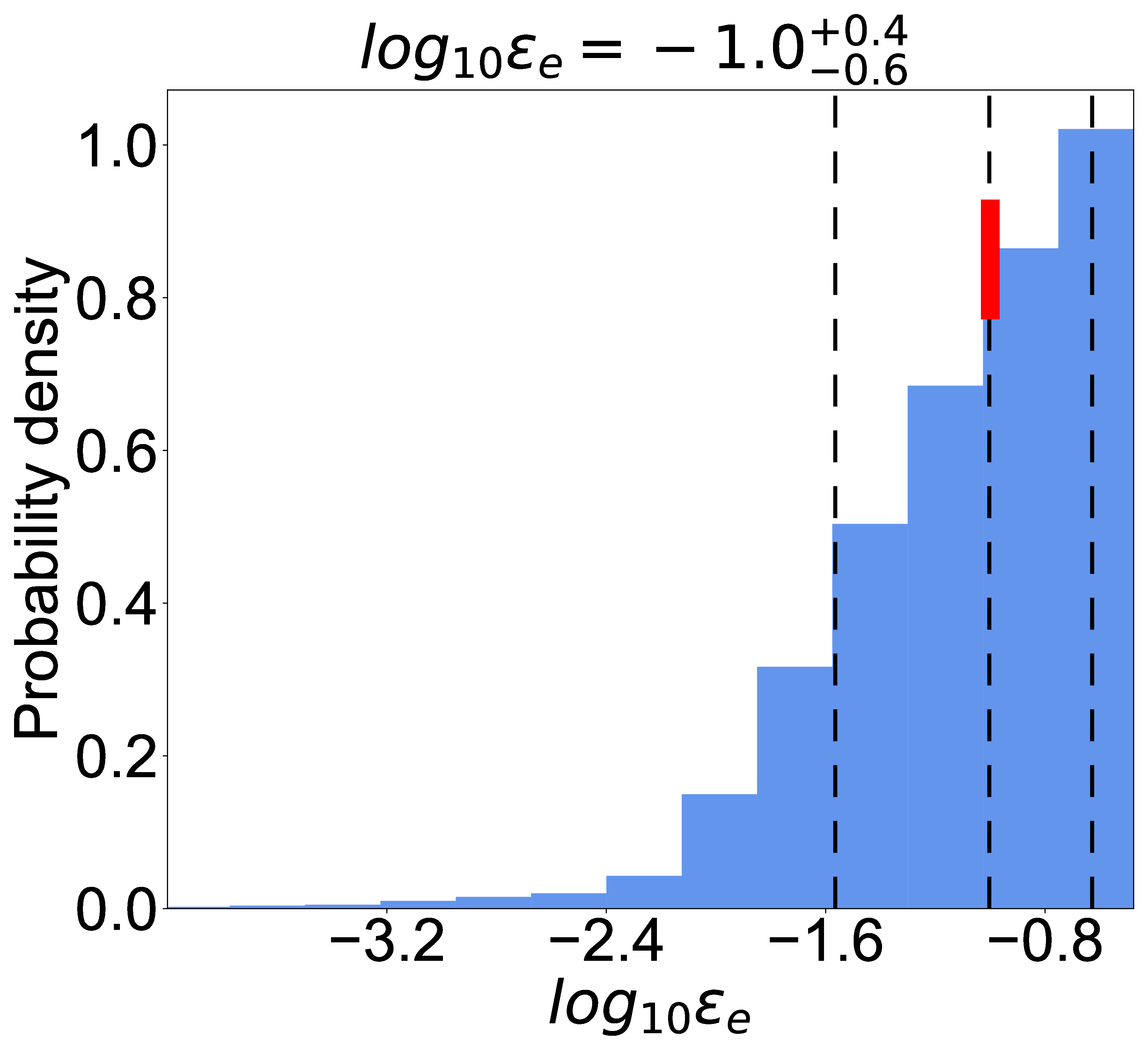}
\hspace{0.3cm}
\includegraphics[height=4cm,width=4cm,trim=4 7 8 12, clip]{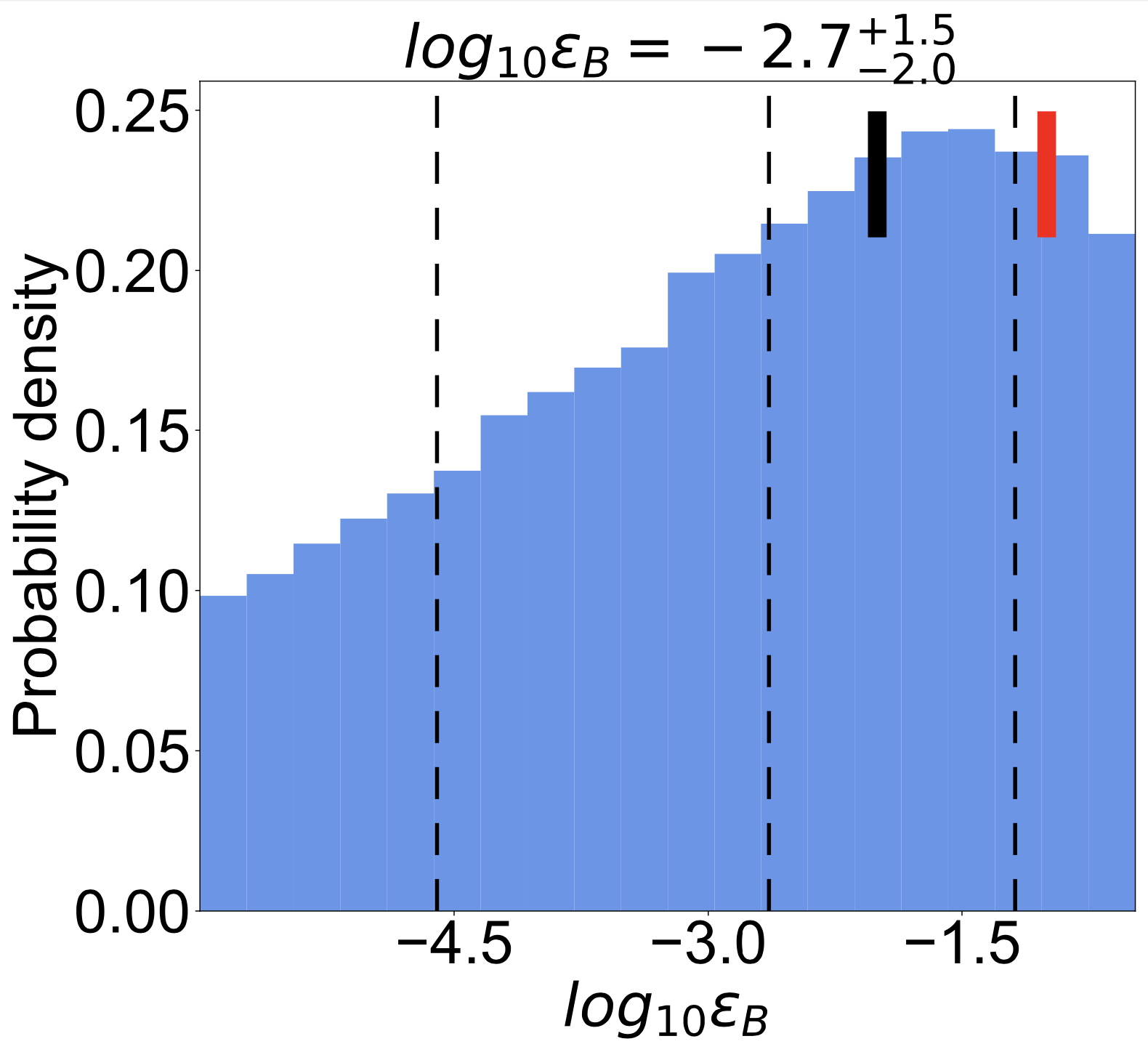}
\begin{center}
\includegraphics[height=4cm,width=4cm,trim=4 7 8 12, clip]{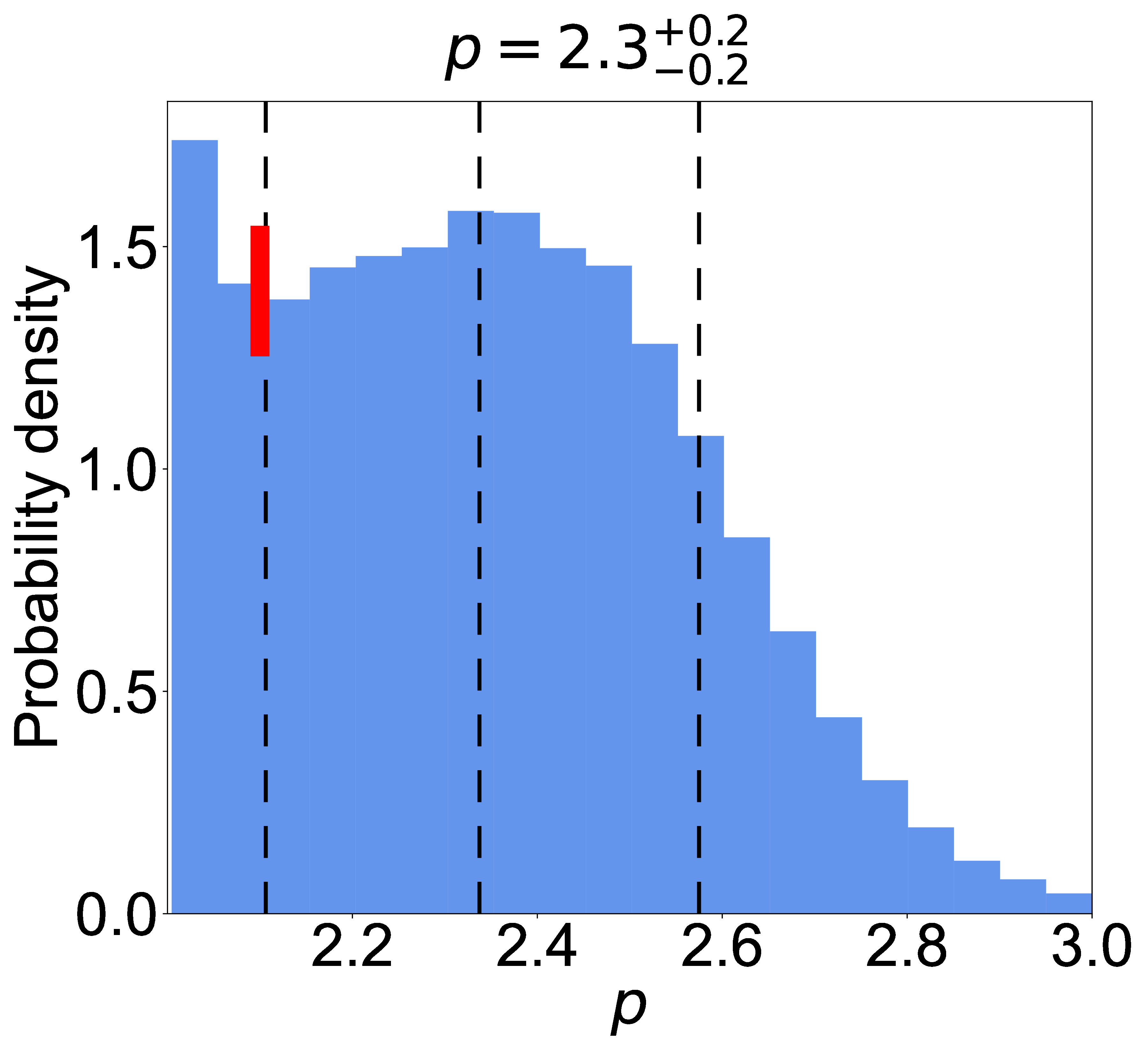}
\end{center}
\caption{ Posterior distributions of the afterglow best fit parameters for GRB 181123B (same as Figure~\ref{afterglow-fit}) obtained taking into account Synchrotron Self-Compton (SSC) cooling in the Thomson regime (i.e. without applying Klein-Nishina corrections).
}
\label{afterglow-fit_SSC}
\end{figure}

\end{document}